\documentclass{aa}  

\usepackage{graphicx}
\usepackage[varg]{newtxmath}
\usepackage{newtxtext}
\usepackage{booktabs}
\usepackage{xcolor}
\usepackage{lscape}
\begin{document}

   \title{Testing the disk instability model of cataclysmic variables}
\subtitle{Are dwarf novae too bright to be DIM?}
   \author{Guillaume Dubus
		\inst{1,3}
		\and
	   	Magdalena Otulakowska-Hypka
		\inst{2}
		\and
		Jean-Pierre Lasota
		\inst{3,4}
          }

   \institute{
   	Univ. Grenoble Alpes, CNRS, IPAG, 38000 Grenoble, France 
         \and
	Astronomical Observatory Institute, Faculty of Physics, Adam Mickiewicz University, S\l{}oneczna 36, PL-60-286 Pozna\'n, Poland
	\and
	Institut d'Astrophysique de Paris, CNRS and Sorbonne Universit\'e, UMR 7095, 98bis Bd Arago, 75014 Paris, France 
	\and
	Nicolaus Copernicus Astronomical Center, Polish Academy of Sciences, Bartycka 18, 00-716 Warsaw, Poland 
	     }

   \date{Received ; accepted ; in original form \today}

 
  \abstract
   {The disk instability model attributes the outbursts of dwarf novae to a thermal-viscous instability of their accretion disk, an instability to which nova-like stars are not subject.}
   {We aim to test the fundamental prediction of the disk instability model: the separation of cataclysmic variables (CVs) into nova-likes and dwarf novae depending on orbital period and mass transfer rate from the companion.}
   {We analyse the lightcurves from a sample of $\approx$ 130 CVs with a parallax distance in the {\it Gaia} DR2 catalogue to derive their average mass transfer rate. The method for converting  optical magnitude to mass accretion rate is validated against theoretical lightcurves of dwarf novae.}
   {Dwarf novae (resp. nova-likes) are consistently placed in the unstable (resp. stable) region of the orbital period -- mass transfer rate plane predicted by the disk instability model. None of the analysed systems present a challenge to the model. These results are robust against the possible sources of error and bias that we investigated. Lightcurves from {\it Kepler} or, in the future, the LSST or {\it Plato} surveys, could alleviate a major source of uncertainty, the irregular sampling rate of the lightcurves, assuming good constraints can be set on the orbital parameters of the CVs that they happen to target.}
   {The disk instability model remains the solid base on which to construct the understanding of accretion processes in cataclysmic variables.}
\keywords{accretion, accretion disks -- binaries: close -- stars: dwarf novae -- novae, cataclysmic variables}

   \maketitle
%

\section{Introduction}
Cataclysmic variables (CVs) are binary systems composed of a white dwarf accreting from a low-mass stellar companion filling its Roche lobe \citep{Warner:1995mo}. CVs come in an bewildering variety of types and subtypes corresponding to their variability properties and spectra. The overarching distinction is between the dwarf novae, which show repeated outbursts with an amplitude $\ga 2$ optical magnitudes on timescales of weeks to decades, and the nova-like systems, whose lightcurves remain roughly steady on the same timescales. 

The outbursts of dwarf novae have long been known to originate in the accretion disk surrounding the white dwarf \citep{1971AcA....21...15S,Osaki:1974uz}, due to a mechanism identified by \citet{Meyer:1981ve}. This instability occurs when the temperature is low enough in the accretion disk that hydrogen recombines. The steep dependence of the opacity with temperature in this regime triggers a thermal and a viscous instability that leads the disk to cycle through two states. In the eruptive state, the disk has a high temperature $>10^4\,$K, hydrogen is highly-ionised, and the mass accretion rate $\dot{M}$ is higher than the mass transfer rate $\dot{M}_t$ from the companion star. In the quiescent state, the disk has a temperature $\la 3000\,$K, hydrogen is mostly neutral, and $\dot{M}< \dot{M}_t$. The {\it disk instability model} (DIM) aims at exploring the consequences of this instability on disk accretion and explaining the variety of observed lightcurves \citep{1996PASP..108...39O,Lasota:2001th}.

A fundamental test of the DIM is whether it reproduces the distinction between dwarf novae and nova-likes. In nova-likes, the whole disk should be hot enough for hydrogen to be highly-ionised. This translates into a minimum mass accretion rate above which a disk of given size is stable. The system is a dwarf novae if this criterion is not met. Hence, the DIM can be tested against observations by deriving the average mass accretion rate  and the size of the disk $R_{\rm out}$ in dwarf novae and nova-likes. Importantly, this  instability criterion does not depend  on the details of how matter is transported in accretion disks, which remains a highly-debated issue.

\citet{1982AcA....32..213S} showed that this criterion roughly separates nova-likes from dwarf novae in the $(R_{\rm out}, \dot{M})$ plane, using approximate magnitudes and distances for a dozen CVs. Since then, only \citet{2007A&A...473..897S} have updated this work, for a sample of 10 CVs. The importance of this test was revived by \citet{2002AA...382..124S}, who pointed out that the {\it HST} FGS parallax distance of SS Cyg, the archetypal dwarf nova, led to a revised mass transfer rate that placed this outbursting system in the {\it stable} region of the   $(R_{\rm out}, \dot{M})$ plane.  This led to much hand-wringing \citep{2007A&A...473..897S,2010AcA....60...83S},  until a radio VLBI parallax distance \citep{2013Sci...340..950M}, confirmed with {\it Gaia} \citep{2017A&A...604A.107R}, firmly placed the system in the unstable region \citep[as anticipated by][]{Lasota2008} .

Although there are dozens of CVs with relatively well-known binary parameters, a thorough investigation of this test has yet to be performed.  \citet{2017A&A...604A.107R} proposed to test the DIM by taking the absolute magnitude at a specific stage of a dwarf nova outbursts (at the end of long outbursts or in standstills) to compare the derived mass accretion rate to the critical rate for the system, under the assumption that both should be nearly equal. Good agreement was found for the sample of a dozen CVs with distances in the {\it Gaia} first data release. Its main advantage is that the required optical magnitude is straightforward to extract from the lightcurves. However, a major drawback of this approach is that it relies on assumptions on how matter is transported in the disk, because one needs to predict when the mass accretion rate will happen to  be equal to the critical rate during the outburst cycle. Hence, it is a much less fundamental and robust test of the basic premises of the DIM. 

Indeed, an important difficulty in using the fundamental test is that it requires extracting the average mass transfer rate from CV optical lightcurves which in many cases are unevenly, inhomogeneously and incompletely sampled. The situation is somewhat easier for  X-ray binaries, where the DIM also applies, as X-ray monitors provide  continuous coverage and X-rays provide a closer measure of the bolometric luminosity in those systems than optical magnitudes do for CVs, which in outburst emit primarily UV radiation. \citet{2012MNRAS.424.1991C} were able to derive average mass accretion rates for $\sim$ 50 X-ray binaries and show that the DIM, modified to include X-ray irradiation \citep{van-Paradijs:1996dz}, separates consistently transients from steady systems.

Here, we carry out, for the first time on CVs, a similar analysis to that performed on X-ray binaries by  \citet{2012MNRAS.424.1991C} in order to test the fundamental prediction of the disk instability model: the separation of cataclysmic variables into stable and unstable systems is determined by the orbital period (a proxy for disk size) of the binary and mass transfer rate from the companion. We use a sub-sample of the CV lightcurves gathered by \citet{2016MNRAS.460.2526O}. We consider all types of dwarf-nova outbursts, including  so-called \textsl{super-outbursts} that are not supposed to be described by the standard version of the DIM \citep[see, e.g.,][and references therein]{Lasota:2001th} but which have the same instability criterion. We first validate the conversion from optical magnitudes to accretion rates against model lightcurves (\S2). The sample of CVs and our approach to analysing their lightcurves is described in \S3. \S4 discusses the results and  possible sources of errors. We conclude on the validity of the DIM.

\section{Model\label{s:model}}

\subsection{Converting disk flux to magnitudes}
The accretion disk in CVs is geometrically thin and radiatively efficient. We model the disk emission as the sum of local blackbody spectra at the local effective temperature $T_{\rm eff}$, which is a function of radius $R$ and time. The flux is
\begin{equation}
F_\nu=\frac{\cos i}{d^2} \int_{R_{\rm in}}^{R_{\rm out}} B_\nu 2\pi R   dR
\label{eq:flat}
\end{equation}
with
\begin{equation}
B_\nu=\frac{2h\nu^3}{c^2}\left[\exp\left(\frac{h\nu}{k T_{\rm eff}}\right)-1\right]^{-1}
\end{equation}
$i$ the disk inclination and $d$ the distance to the binary.  The blackbody approximation is good when compared to using stellar spectra \citep{1989AcA....39..317S}. The disk instability model provides the evolution of $T_{\rm eff}$ as a function of radius and time. The disk emission is converted to an optical flux in the Johnson $V$ band  
\begin{equation}
F_V=\frac{\cos i}{d^2} \int_{R_{\rm in}}^{R_{\rm out}} w_V(T_{\rm eff})\sigma T_{\rm eff}^4   2 \pi R  dR
\label{eq:diskflux}
\end{equation}
with $w_V(T)$ defined as   
\begin{equation}
w_V(T)=\frac{1}{\sigma T^4}\frac{\int f_V B_\nu d\nu}{\int f_V d\nu}
\label{eq:bolcor}
\end{equation}
$f_V$ is the filter response from \citet{2015PASP..127..102M}, as implemented in  the SVO Filter Profile Service \footnote{The SVO Filter Profile Service. Rodrigo, C., Solano, E., Bayo, A. \url{http://ivoa.net/documents/Notes/SVOFPS/index.html} and the Filter Profile Service Access Protocol. Rodrigo, C., Solano, E. \url{http://ivoa.net/documents/Notes/SVOFPSDAL/index.html}}. The absolute magnitude is then given by
\begin{equation}
M_V=-2.5 \log\left[ \left(\frac{d}{\rm10\,pc}\right)^2 \left(\frac{F_V}{Z_V}\right)\right]
\label{eq:absmag}
\end{equation}
where $Z_V=3562.5$\,Jy is the $V$ band flux zeropoint in the Vega system.  Numerical integration of Eq.~\ref{eq:diskflux} for a given radial distribution of $T_{\rm eff}$ in the disk, using the bolometric correction (Eq.~\ref{eq:bolcor}) gives the disk's absolute magnitude $M_V$ (Eq.~\ref{eq:absmag}). The same procedure can be used for any filter in the database.

\subsection{Disk parameters}
The ratio of outer disk size to binary separation $R_{\rm out}/a$ depends only on the mass ratio $q$. Using \citet{1977ApJ...216..822P} or \citet{1979MNRAS.186..799L} gives comparable results. The typical disk size is between (2 to 7)$\times 10^{10}$\,cm for $80{\rm\,mn}\leq P_{\rm orb}\leq 10$\,hrs. The disk circularisation radius in units of binary separation $a$, $R_{\rm circ}/a$, also depends only on $q$.

The minimum inner disk radius is set to the white dwarf radius, parametrised as \citep{1972ApJ...175..417N}
\begin{equation}
R_{\rm WD}=7.8\times 10^8\ \left[\left(\frac{1.44\rm\,M_\odot}{M}\right)^{2/3}-\left(\frac{M}{1.44\rm\,M_\odot}\right)^{2/3}\right]^{1/2}{\rm\,cm}
\label{eq:wdradius}
\end{equation}
However, the magnetic field of the white dwarf can truncate the inner disk before it reaches the white dwarf surface. Following the standard approach \citep{Frank:2002fo}, the truncation radius is set by
\begin{equation}
R_{\rm mag}\approx\left(\frac{\mu^4 }{GM\dot{M}_{\rm in}^2} \right)^{1/7}
\label{eq:trunc}
\end{equation}
 with $\mu=B_{\rm WD} R_{\rm WD}^3$ the magnetic moment of the white dwarf and $\dot{M}_{\rm in}$ the mass accretion rate at the inner radius. The truncation radius is close to or at the white dwarf radius in outburst, when $\dot{M}_{\rm in}$ is high. It can be much larger than the white dwarf radius in quiescence. Besides changing the outburst properties, this has two main consequences for the present work: (1) a cold stable disk can exist if the truncation radius is large; (2) the varying inner disk radius changes the relationship between magnitude and mass transfer rate. 
 
Limb darkening changes the inclination dependence compared to the simple $\cos i$ dependence (Eq.~\ref{eq:flat}). We obtain the inclination dependence by using the magnitude correction derived from more elaborate disk spectral models by  \citet{1980AcA....30..127P}
\begin{equation}
\Delta M=-2.5 \log \frac{F(i)}{\left<F\right>}=-2.5\log\left[ \left(1+\frac{3}{2}\cos i\right)\cos i \right]
\end{equation}
where $\left<F\right>=F(i=0\degr)/2$ is the  flux averaged over all inclinations. This works well for $i\la75\degr$.

\subsection{The $\dot{M}$-$M_V$ relationship during a dwarf nova cycle}
\begin{figure}
\begin{center}
\includegraphics[width=0.9\linewidth]{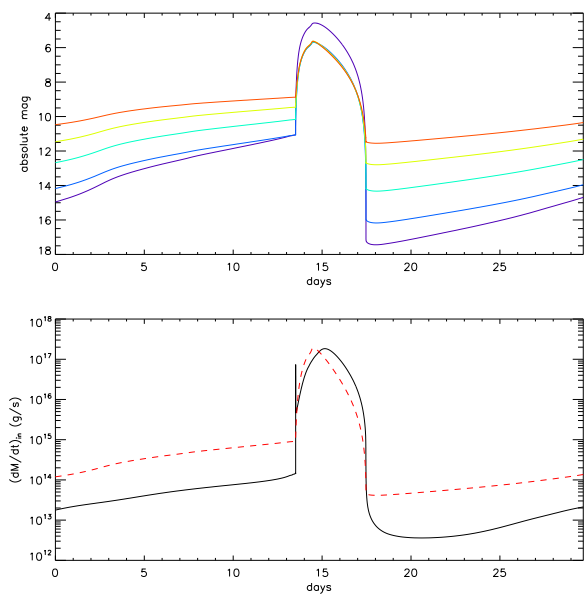} 
\caption{Absolute magnitude in $U$, $B$, $V$, $R$, $I$ bands (blue to red, top panel) and mass accretion rate onto the white dwarf $\dot{M}(R_{\rm in})$ during a dwarf nova outburst. The model parameters are $M=0.6\rm\,M_\odot$, $\dot{M}_{\rm t}=10^{16}\rm\,g\,s^{-1}$, $R_{\rm out}\approx 2\times 10^{10}\rm\,cm$, $i=0\degr$. The dashed line shows the  $\dot{M}(R_{\rm in})$ reconstructed from the $V$ magnitude lightcurve assuming $\dot{M}$ does not depend on radius. The reconstructed mass transfer rate is 9$\times 10^{15}\rm\,g\,s^{-1}$ after averaging over the outburst cycle.}
\label{fig:light}
\end{center}
\end{figure}

The goal is to reconstruct the mass transfer rate from the optical lightcurve. Assuming no mass is lost from the disk except by accretion onto the white dwarf, then the mass transfer rate is equal to the time average of the mass accretion rate $\dot{M}(R,t)$ through the disk. Ideally, we would use the bolometric luminosity of the disk to reconstruct $\dot{M}_{\rm in}$. This is not feasible because of lack of multi-wavelength data covering UV (where most of the accretion luminosity is emitted) and/or sufficient time coverage of the outbursts. We must therefore rely on optical magnitudes as a function of time, typically $V$  (Fig.~\ref{fig:light}).

\begin{figure}
\begin{center}
\includegraphics[width=0.9\linewidth]{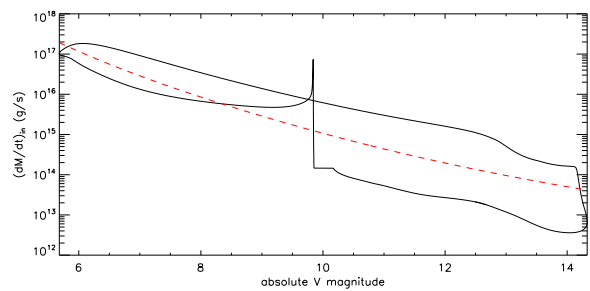} 
\caption{Relation between the absolute $V$ magnitude and the inner mass accretion rate $\dot{M}_{\rm in}$ for the model shown in Fig.~\ref{fig:light}. The disk samples the black curve clockwise during the outburst cycle. The dashed line shows the relationship using the stationary disk approximation.}
\label{fig:mvmdot}
\end{center}
\end{figure}

The radial distribution of $T_{\rm eff}$ varies in a complex fashion during the dwarf nova cycle as fronts propagate through the disk (see e.g. \citealt{1998MNRAS.298.1048H}). Schematically, $T_{\rm eff}$ changes from a steady-state like $R^{-3/4}$ profile in outburst to a nearly flat profile in quiescence. The result is a hysteresis between the optical magnitude and the mass accretion rate  (Fig.~\ref{fig:mvmdot}).  Ideally, we would build a dwarf nova model for each system so that we can precisely obtain the conversion factor from magnitude to $\dot{M}_{\rm in}$ (Fig.~\ref{fig:mvmdot}). We have used a much simpler method, at the price of a modest error in accuracy. 

In the following, we assume that $\dot{M}$ does not depend on radius {\it i.e.} that the evolution behaves as a succession of steady states with $\dot{M}=\dot{M}_{\rm in}$. This is not correct, as discussed above, but, as we shall prove, the corresponding error is entirely acceptable given the other  uncertainties. For a stationary disk \citep{Frank:2002fo}, 
\begin{equation}
\sigma T_{\rm eff}^4=\frac{3}{8\pi}\frac{GM\dot{M}}{R^3}\left[1-\left(\frac{R_{\rm in}}{R}\right)^{1/2}\right]
\end{equation}
so that the absolute magnitude depends only on mass accretion rate $\dot{M}$, white dwarf mass $M$, inner disk radius $R_{\rm in}$ and outer disk radius $R_{\rm out}$. The integration to get $M_V$ for given $(\dot{M},M,R_{\rm in},R_{\rm out})$ is straightforward. We checked that our routine gives $\dot{M}-M_V$ relationships that are identical to those shown in \citet{1980AcA....30..127P} and \citet{1989AcA....39..317S}, who used the same assumptions, and to the $M_V$ values found by \citet{1981AcA....31..127T}, who used a more elaborate spectral model. The relationship can be inverted to give $\dot{M}$ as a function of $M_V$ although the rightmost term in the expression of $T_{\rm eff}$ (related to the no-torque inner boundary condition) requires solving an equation by iteration.  The $\dot{M}$ inferred from the absolute $V$ magnitude is shown as dashed red lines in Fig.~\ref{fig:light}-\ref{fig:mvmdot}. As we verify below, averaging this $\dot{M}$ over an outburst cycle gives a reconstructed mass transfer rate that is close to the true input mass transfer rate $\dot{M}_{\rm t}$. 

\subsection{Validation against model lightcurves\label{s:validation}}
The error made in using the steady state  approach can be quantified by using model disk lightcurves with parameters covering a range of possible values for CVs. Optical lightcurves ($U$, $B$, $V$, $R$, $I$) were obtained by following the evolution of the disk with the code described in \citet{1998MNRAS.298.1048H}. Table~\ref{table:model} lists the parameters used for the models. The size of the disk is free to vary during the outburst as the angular momentum from the disk is removed by tidal forces from the companion. The inner disk radius is kept fixed except in the last six models where it is left free to vary according to magnetic truncation with $\mu\sim10^{30}\rm\,G\,cm^{3}$.

The input mass transfer rate $\dot{M}_{\rm t}$ was compared to the reconstructed $\dot{M}_{\rm filter}$ from the various optical lightcurves, averaged over an outburst cycle. The reconstruction used the average outer disk radius, the exact value of the white dwarf mass, and set $R_{\rm in}$ to the value given by Eq.~\ref{eq:wdradius} (even when the inner disk radius varied due to magnetic truncation). Inclination effects were not considered here. The reconstructed rate nearly always underestimate $\dot{M}_{\rm t}$, all the more so that the filter is redder, but not by much: for example the average $\dot{M}_V/\dot{M}_{\rm t}=0.85\pm0.10$ for the models in Tab.~\ref{table:model}. This bias, due to the assumptions of the model, is negligible compared to the other sources of error, notably the inclination (\S\ref{s:error}).

\section{Optical lightcurves}

\subsection{Source list and system parameters\label{s:param}}
For the dwarf novae, we analysed a subset of the lightcurves from the dataset collected and presented in \citet{2016MNRAS.460.2526O}. We required a value for $P_{\rm orb}$, a parallax distance $d$, and at least a 5 year timespan of observations to include it in the analysis. The parallax $\pi$ and its associated uncertainty were taken from the {\it Gaia Data Release 2} \citep{2016A&A...595A...1G,2018arXiv180409365G,2018arXiv180409366L}. The  superb quality of the {\em Gaia}-based distances remove what was previously the major source of uncertainty in deriving the mass transfer rate. Table~\ref{table:DN} lists the dwarf novae systems and the parameters that we adopted. The same approach was followed for nova-likes (Table~\ref{table:NL}), selecting systems with a  {\it Gaia} parallax and using  lightcurves from the AAVSO database \citep{aavso}. AM Her systems (polars) were excluded as there is no disk in those systems, where accretion is entirely controlled by the high magnetic field of the white dwarf. The values for $q=M_1/M_2$, $M_1$, $M_2$, $i$, and the extinction $A_V$ have been mined from a variety of sources (see the reference list below Tab.~\ref{table:DN}). In some cases these values may vary significantly for the same source; in many other we did not find an estimated value in the literature  and we had to supplement those as described below. When several values were available, we generally kept the value from the latest reference. 

Italics in Tab.~\ref{table:DN}-\ref{table:NL}  distinguish the values that we assumed, or derived from other parameters, from those that we took from the literature. If only $M_1$ or $M_2$ were available we used the mass ratio $q$ to deduce the other. When neither was available, we estimated $M_2$ from the evolutionary sequence with $P_{\rm orb}$ of \citet{2011ApJS..194...28K}, taking an estimated relative error $\Delta M_2/ M_2=0.2$ and deducing $M_1$ from $q$. When $q$ was unavailable, we adopted $M_1=0.75\rm\,M_\odot$ for consistency with \citet{2011ApJS..194...28K}. The error on $M_1$ and $M_2$ was propagated from the error on $q$ wherever necessary.  We only selected systems with $i\leq 80\degr$ as the magnitude correction is incorrect for edge-on systems. When we could not find an estimated value for $i$, we took $i=56.7\degr\pm20\degr$: this value of $i$ corresponding to zero magnitude correction. We took $\Delta i=\pm10\degr$  when no error was available on the system inclination $i$. The median extinction $A_{V}$ at the location and parallax distance of each object was obtained from the 3D map of Galactic reddening by \citet{2018arXiv180103555G}. A rough error was estimated by taking the maximum difference between the median $A_V$ and the 16\% or 84\% percentile (i.e. the 1$\sigma$ interval).  We used values from the other references listed below Tab.~\ref{table:DN} for the few binaries whose location is not covered by this map. We did not find values for ST Cha and AT Ara: we took the $A_V$ along the line-of-sight from \citet{2011ApJ...737..103S}. When no error was available we took $\Delta A_V=0.1$. We note that the 3D reddening map of \citet{2018arXiv180103555G} gives a range $A_{V}=0.010$ (16\% percentile) to $0.063$ (84\% percentile) in the direction and at the distance of SS Cyg, consistent with the $A_{V}=0.062\pm0.016$ specifically derived for this important source by  \citet{2013PASP..125.1429R}. We used the 3D map median value $A_{V}=0.017$.

\subsection{Lightcurve analysis}
Ideally, the lightcurves should be regularly sampled on sub-orbital timescales and cover several outburst cycles for the dwarf novae. However, most lightcurves  are irregularly sampled with, for example, the frequency of observations increasing during outbursts or during dedicated campaigns. The lightcurves may also feature orbital modulations or eclipses. The data must be homogenised in time to derive the time-averaged optical magnitude of the binary. To do this, we averaged the data over a 1 day sliding window, incrementing by 1 hour between each bin.  This smoothes over orbital modulations without affecting the rise times and decay times to outburst. To avoid  large gaps with missing data, we  manually selected the portion of the lightcurve that we deemed long enough to average over multiple outburst cycles while remaining densely-sampled. The start and stop days we chose are listed in Tab.~\ref{table:DN}. 
 
We define the filling fraction $f$ of the resulting lightcurve as the fraction of its bins that contain a measurement. For the dwarf novae, we assumed the  missing measurements corresponded to zero flux. We verified that assuming that the flux is equal to an observationally-defined non-zero quiescent value makes no difference to our results. However, we may be missing outbursts, which would have a larger impact on the estimated mass accretion rate. With this in mind, we separated our dataset between lightcurves that have $f>0.5$, minimising the impact of missing data, and lightcurves with $f<0.5$, where the uneven sampling causes a larger uncertainty.  We discuss the bias this introduces below (\S\ref{s:error}). 

For the nova-likes, where the source variability is more limited, we assumed that the missing measurements were at the average flux seen at other times. Hence, the filling fraction $f$ carries much less weight for these systems. Otherwise, we followed the same analysis as dwarf novae. The time interval over which we analyse the lightcurves are indicated in Tab.~\ref{table:NL}: for VY Scl systems we took the flux during steady maxima of the lightcurve. Table~\ref{table:NL} also lists the mean $V$ magnitude of the selected portion of the lightcurve in order to ease comparison with values in the literature for those steady systems.

The magnitudes were then corrected for extinction, distance, and disk inclination (taken to be in the binary plane). We did not correct for contributions to the $V$ band flux from the bright spot, companion star, and white dwarf, as we discuss below (\S\ref{s:error}). The absolute magnitudes were converted to mass accretion rates, following the procedure described in \S\ref{s:model}, and time-averaged to obtain the mass transfer rate.

We calculate the statistical error on the mass transfer rate through Monte Carlo sampling of the errors on the mass of the binary components, inclination, parallax and $A_V$. The parallax and extinction probability distributions are taken to be Gaussians whereas the mass distributions and $\cos i$ are sampled uniformly. Our statistical error estimate is the smallest 68\% confidence interval from the distribution of Monte Carlo sampled mass transfer rates, which is close to a log-normal.

\section{Results}

\subsection{The critical mass transfer rate}

\begin{figure*}
\begin{center}
\includegraphics[width=\linewidth]{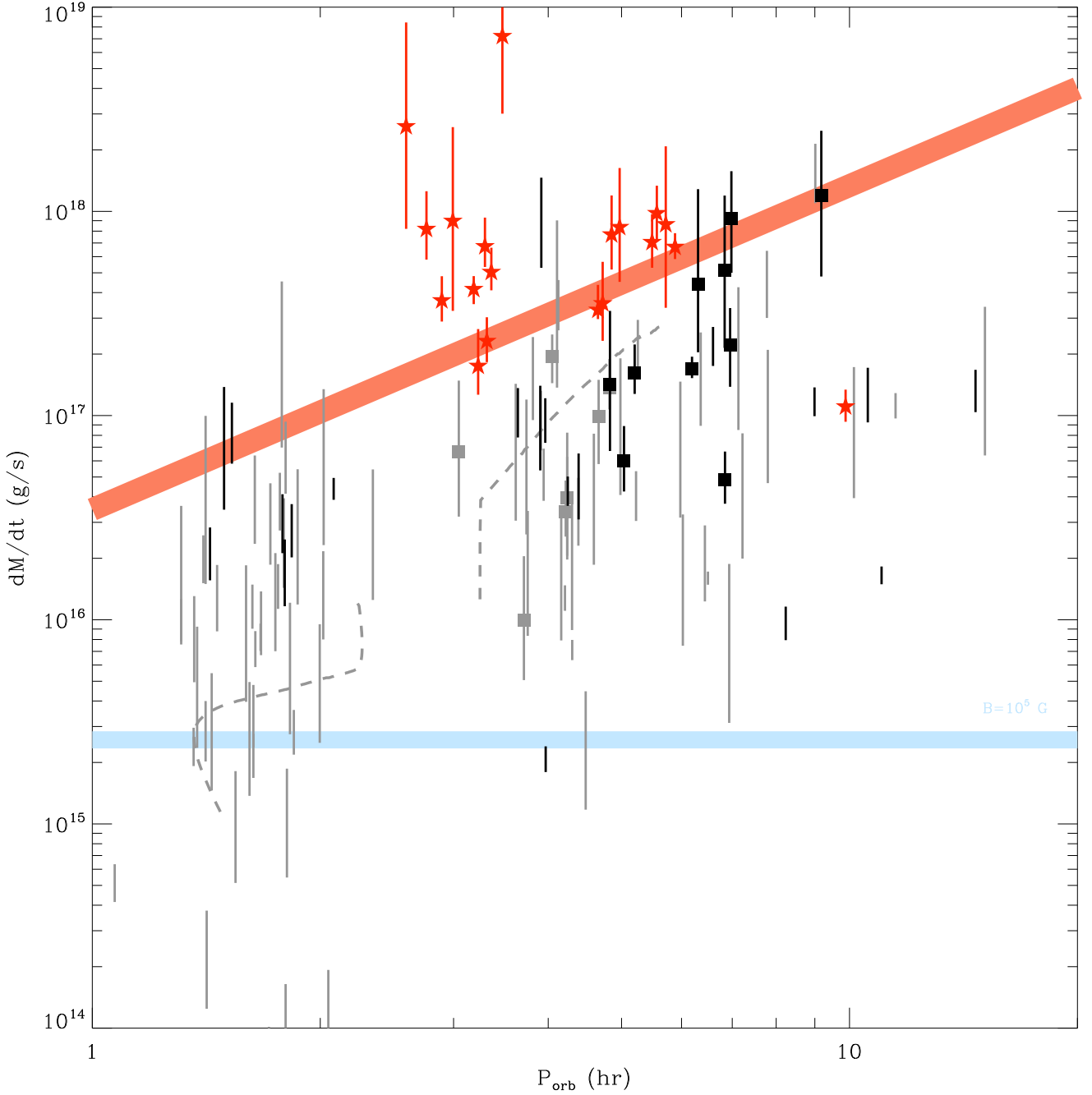} 
\caption{Mass transfer rates of cataclysmic variables compared to the stability criterion. Systems above the (red) upper solid line are hot and stable. Systems below the lower (blue) line will be cold, stable disks if the white dwarf magnetic field $B\geq 10^5$\,G. Dashed line is the expected secular mass transfer rate \citep{2011ApJS..194...28K}. Square symbols indicate Z Cam type dwarf novae, (red) stars indicate nova-likes. Dwarf novae with $f\geq 0.5$ are shown in black and those with $f<0.5$ are in grey. Fig.~\ref{fig:stability_names} in the Appendix includes system names.}
\label{fig:stability}
\end{center}
\end{figure*}
Figure~\ref{fig:stability} compares the calculated mass transfer rates against the critical mass transfer rate (Fig.~\ref{fig:stability_names} includes the system names to help localise them in the diagram). The critical mass transfer rate above which a cataclysmic variable with a disk size $R$ is hot and stable is 
\begin{equation}
\dot{M}_{\rm crit}^{+}=8.07\times10^{15}~R_{10}^{2.64}~M_1^{-0.89}\,\rm g\,s^{-1}
\label{eq:mdot+}
\end{equation}
using the fit for a disk with solar composition given in \citet{2008A&A...486..523L}, with $R_{10}=R/ 10^{10}\rm\,cm$ and $M_1$ the white dwarf mass in solar masses. Similarly, the critical mass transfer rate below which a CV with an {\it inner} disk radius $R$ is cold and stable is 
\begin{equation}
\dot{M}_{\rm crit}^{-}=2.64\times10^{15}~R_{10}^{ 2.58}~M_1^{-0.85}\,\rm g\,s^{-1}
\end{equation}
This is usually too low to be of consequence if the disk extends to the white dwarf surface (low R). However,  if the inner disk radius is truncated by the white dwarf magnetic field then, combining with Eq.~\ref{eq:trunc}
\begin{equation}
\dot{M}_{\rm crit}^{-}=8.78\times10^{13}~\mu_{30}^{ 0.85}~M_1^{-0.70}\,\rm g\,s^{-1}
\label{eq:mdot-}
\end{equation}
with $\mu_{30}=\mu/10^{30}\rm\,G\,cm^3$ the magnetic moment of the white dwarf. A reasonably strong magnetic field can thus truncate the disk and keep it cold and stable if the mass transfer rate is low. This can be used to place an upper limit on the magnetic field of the white dwarf in a dwarf nova. 

Figure~\ref{fig:stability} shows linear fits to the values of $\dot{M}_{\rm crit}^{+}$ and $\dot{M}_{\rm crit}^{-}$ obtained for each object, with the width of the line covering the range of individual values. The plotted $\dot{M}_{\rm crit}^{-}$ corresponds to  a white dwarf magnetic field $B=10^5\rm\,G$ or $\mu\approx 3\times 10^{32}\rm\,G\,cm^{3}$, which is the typical magnetic field above which cataclysmic variables are identified as intermediate polars (DQ Her type).  For $\dot{M}_{\rm crit}^+$, we have 
\begin{equation}
\dot{M}_{\rm crit}^{+}\approx 3.5\times 10^{16}\ P_{\rm orb}^{1.6}\rm\,g\,s^{-1}
\end{equation}
with  $P_{\rm orb}$ in hours. This is roughly recovered by noting that $R_{\rm out}\propto a \propto P_{\rm orb}^{2/3} M_1^{1/3}$ up to some slowly-varying function of $q$. This also shows that the dependence of $\dot{M}_{\rm crit}^+$ on $M_1$ basically cancels out when $R_{\rm out}$  is replaced by $P_{\rm orb}$ in Eq.~\ref{eq:mdot+}. Except for $P_{\rm orb}$, which is usually precisely known, the errors in system parameters have very little influence on $\dot{M}_{\rm crit}^{+}$.
 
\subsection{Dwarf novae}
All the dwarf novae are well within or consistent with the unstable region delimited by  $\dot{M}_{\rm crit}^{+}$, clearly confirming a key prediction of the disk instability model. There are 8 systems where the estimated $\dot{M}_t$  is above the stability limit. However, for 7 of them, the estimated $\dot{M}_t$ is still compatible with the unstable region within the statistical uncertainty (AT Ara, EM Cyg, ER UMa, SY Snc, V1159 Ori,  FO Per, V516 Cyg, V1316 Cyg). CN Ori is the only system where the estimated error bar is inconsistent with the critical transfer rate: it has an estimated $\dot{M}_t=(5.4\,{\rm to}\, 14.8) \times 10^{17}\rm\,g\,s^{-1}$ to compare to $\dot{M}_{\rm crit}^{+}=2.7\times 10^{17} {\rm\,g\,s^{-1}}$. A likely reason for this discrepancy is the  median $A_V=0.48$ provided by the 3D reddening map, which is flagged as unreliable for such a nearby source. The extinction could be much lower: indeed the "best fit" value  from the model of \citet{2018arXiv180103555G} is $A_V=0.083$ and \citet{2006ApJ...642.1029U} list $A_V\approx 0$. The estimated $\dot{M}_t$ becomes compatible with $\dot{M}_{\rm crit}^{+}$ if there is negligible extinction. The error on the inclination, $i=67\degr\pm3\degr$, may also be underestimated. Given this and the level of the systematic uncertainty on our measurements (see \S\ref{s:error}), we conclude that none of the systems are incompatible with the stability limit.  

SU UMa stars, dwarf novae with superoutbursts, are concentrated below the period gap, as usual, except for VW Vul and ES Dra, which are also classified as Z Cam types in \citet{2014JAVSO..42..177S}. Their classification as SU UMa types in \citet{2001PASP..113..764D} is likely to be incorrect.

Z Cam types tend to have mass-transfer rates higher than other dwarf novae. Their lightcurves have standstills that have long been argued to be due to fluctuations of the mass transfer rate bringing it above the stability limit. One would thus expect Z Cam systems to be close to the stability limit \citep[][]{1983A&A...121...29M,2001A&A...369..925B}. Indeed, fluctuations by a factor 3 would push many of the Z Cam systems very close to or above the stability limit. The lack of Z Cam types below the period gap, even though some systems are very close to the stability limit, suggests that such fluctuations do not occur in these systems, most likely for reasons related to the type of the companion star. Above the period gap, some Z Cam systems are somewhat further away than a factor 3 from the stability limit (PY Per, V426 Oph, WW Cet). The classification of WW Cet as a Z Cam stars is uncertain in \citet{2001PASP..113..764D} but considered certain in \citet{2014JAVSO..42..177S}. The filling fraction of the PY Per lightcurve, $f\approx 0.34$, may lead us to underestimate $\dot{M}_ t$ if the flux always remains at a high level. For instance, assuming that the missing flux values are at the average of the measured values increases $\dot{M}_t$ by a factor $\approx 2$ (\S\ref{s:error}). Its disc inclination is also currently unknown and may be high.  V426 Oph is listed as a Z Cam, but also as a DQ Her type in \citet{2001PASP..113..764D} and as a nova-like system in the GCVS. Disk truncation by the white dwarf field may lead us to underestimate the true $\dot{M}_t$ (see \S\ref{s:nl} below). The nova-like  classification is more puzzling, although {\it downward} fluctuations of the mass transfer rate in V426 Oph might push the system into the cold stable regime with a white dwarf magnetic field $B \approx 10^6\rm\,G$.  Inversely, not all of the systems that are very close to the stability limit in Fig.~\ref{fig:stability} are classified as Z Cam types (e.g. FO Per, CY Lyr, CN Ori) but many U Gem stars could be ``unrecognised Z Cam stars'' \citep{Warner:1995mo}. 

We also find there is considerable variation of $\dot{M}_t$ at given $P_{\rm orb}$. Some of it is undoubtedly due to inaccurate values of the parameters adopted for some of the systems or biases in the lightcurve analysis (see \S\ref{s:error}). Although there is a general trend for higher values of  $\dot{M}_t$ with increasing $P_{\rm orb}$, the values differ significantly from the theoretical secular mass transfer rate derived by \citet{2011ApJS..194...28K}, especially above the period gap. Some of these variations may be due to the mechanism for  angular momentum losses from the system, which drives the secular mass overflow rate from the Roche lobe and is much less certain above the gap than below, where it is mainly due to gravitational radiation. In fact, in Figure~\ref{fig:stability}  the evolutionary tracks correspond to a  version of the angular-momentum-loss mechanisms rescaled to correspond to observations, which leads nevertheless to ``two glaring inconsistencies'' \citep{2011ApJS..194...28K} with the observed distribution of dwarf-novae and nova-likes. Some of these inconsistencies may be due to variations in the mass transfer rate on timescales longer than the outburst cycle but smaller than the secular timescale, such as cycles in the stellar activity.

Finally, we also find that some binaries are well below  $\dot{M}_{\rm crit}^{-}$ (e.g. EI Psc, EZ Lyn),  indicating that the white dwarf magnetic field is constrained by the DIM to $B<10^5\rm\,G$ in those systems (or else they would be cold and stable). 

\subsection{Nova-likes\label{s:nl}}
The nova-likes are clearly at higher mass transfer rates than dwarf novae and are consistently above the DIM stability limit defined by the red line in Fig.~\ref{fig:stability}. Closer inspection reveals six systems in our nova list (Tab.~\ref{table:NL}) with an estimated $\dot{M}_t< \dot{M}_{\rm crit}^+$ (BK Lyn, BG CMi, V603 Aql, IX Vel, UX UMa, AE Aqr). BK Lyn is probably misclassified. It is the only nova-like below the period gap in  \citet{2001PASP..113..764D}  but its recent lightcurve shows variability typical of a ER UMa dwarf nova \citep{2013MNRAS.434.1902P}. We have tagged it as a dwarf nova in Fig.~\ref{fig:stability}. The others are statistically consistent with being in the stable region of the diagram, except for AE Aqr.

AE Aqr is a DQ Her type system well-known for propelling material outside the system due to the very fast rotation of the white dwarf  \citep{1997MNRAS.286..436W}. In this case, we may be severely underestimating the actual mass transfer rate from the secondary because of non-negligible mass loss. There are other systems classified as DQ Her systems (intermediate polars, e.g. BG CMi). The inner disk will be truncated by the white dwarf magnetic field even if it does not rotate fast enough to propel material out of the system as in AE Aqr. We do not take into account this truncation and this will affect our estimated  $\dot{M}_t$ for this type of system. For illustration, consider a steady accretion disk around a $1\rm\,M_\odot$ white dwarf with $\dot{M}=10^{17}\rm\,g\,s^{-1}$ and an outer radius $R_{\rm d}=4\times 10^{10}\rm\,cm$. The absolute magnitude of the disk is $M_V\approx 6.7$ if the inner disk is truncated at $\approx 6\times 10^{9}\rm\,cm$ by a magnetic field $B\approx 10^6\rm\,G$ (Eq.~\ref{eq:trunc}).  The estimated $\dot{M}_t$ will only  be 20\% lower than the actual value, well within our systematic errors, if the disk is assumed to be  truncated at $\approx 10^9\rm\,cm$, corresponding to $B\approx 10^5\rm\,G$ instead of the true value $B= 10^6\rm\,G$. However, the estimated mass transfer rate will be $\approx 2\times 10^{16}\rm\,g\,s^{-1}$ if this magnitude is translated to $\dot{M}$ assuming the inner disk extends to the white dwarf surface at $\approx 5\times 10^8\rm\,cm$. We thus caution that the presence of a magnetic field $B\ga 10^5\rm\,G$ may lead us to underestimate the true mass transfer rate in DQ Her systems, and in particular in BG CMi where $B\approx 4$ to $10\times 10^6\rm\,G$ \citep{1990ApJ...350L..13C}. Finally, we note that TV Col, another DQ Her system, shows infrequent and short outbursts but according to \citet{2017A&A...602A.102H} these cannot be attributed to the disk's thermal-viscous instability.  

\subsection{Sources of error and bias\label{s:error}}
\begin{figure}
\begin{center}
\includegraphics[width=0.8\linewidth]{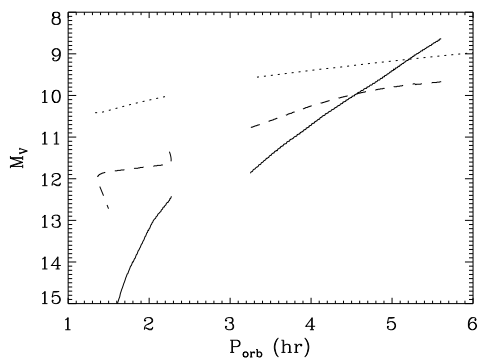} 
\caption{Absolute $V$ magnitude of the companion star (full line) and of the accretion-heated white dwarf (dashed line) as a function of orbital period \citep{2011ApJS..194...28K}. The dotted line is the absolute magnitude of the bright spot, calculated for $\dot{M}=\dot{M}_{\rm crit}$ and a white dwarf mass $M=0.75$\,M$_\odot$.}
\label{fig:cvmag}
\end{center}
\end{figure}
 
The error bars in Fig.~\ref{fig:stability} take into account statistical errors in the system parameters: $M_1$, $M_2$, $i$, $\pi$, $A_V$. The uncertainty on the distance and on the disk inclination dominate the statistical error. However, there are other sources of systematic errors and biases.

First, some of them result from our theoretical assumptions.  The stationary disk assumption in our model biases  $\dot{M}_t$ by $\approx$ 15\% to lower values  (\S\ref{s:validation}). We  also underestimate the true mass transfer rate if the inner disk is truncated by the magnetic field of the white dwarf, as discussed above. We may also underestimate $\dot{M}_t$ if a significant fraction of the mass in the disk is lost to a wind, although estimates of this fraction indicate that the typical wind loss rate represents $\la 10$\% of the mass accretion rate in the hot state \citep[e.g.][]{2010ApJ...719.1932N}. Inversely, the mass transfer rate may be enhanced during the outburst \citep{2000A&A...353..244H,2008AcA....58...55S} due e.g. to the irradiation of the donor star, but this is debated \citep[e.g.][]{2008A&A...489..699V}. We did not compensate for any of these effects. 

Second, the lightcurve analysis also systematically underestimates  $\dot{M}_t$.  To calculate the average mass transfer rate $<$$\dot{M}$$>$, we assume that $\dot{M}=0$ when there are no measurements and mitigate this source of error by taking lightcurves with high filling fractions $f$ i.e. so that outbursts are not missed. Were we instead to assume that $\dot{M}=<$$\dot{M}$$>$ when measurements are missing, the mass transfer rate would be increased to $(2-f)<$$\dot{M}$$>$. This assumption is clearly very unlikely looking at the lightcurves, but taken at face value the mass transfer rates would be underestimated by 50\% with $f=0.5$. A more pernicious error is that caused by high flux outliers in a poorly sampled portion of the lightcurve. A few odd points can thus induce an overestimate of the mass transfer rate. We have minimized this by manually inspecting the lightcurves to remove glaring outliers (a dozen points altogether). We also investigated the influence of our binsize (from 0.25 to 12 hours) and averaging window size (from 0.25 to 2 days). We find differences $\la 50\%$ in $\dot{M}_t$ with no obvious trend as to over- or underestimating $\dot{M}_t$.  

Third, the extinction may be much more uncertain that we assumed, notably for nearby binaries where the reddening measurements of \citet{2018arXiv180103555G} are less reliable for lack of stars or  where we have used the total extinction along the line-of-sight. In general, we have  found good agreement between the values we obtained from the 3D map and those we could collect in the CV literature (e.g. \citealt{2007NewA...12..446A}). Yet, quite different values of $A_{V}$ are sometimes found for the same binary. For instance, RW Tri has $A_V\approx 0.3$ in Tab.~\ref{table:NL} and in \citet{2017PASP..129a2001B} but is listed with $A_V\approx 0.8$ in \citet{2007NewA...12..446A} and with $A_V=0.3 - 0.7$ in \citet{2007A&A...473..897S}. This uncertainty can affect $\dot{M}_{t}$ for some systems but, in our experience, has no impact on the conclusions.

Finally, we may have overestimated the mass transfer rate by spuriously including contributions from the white dwarf, the bright spot and the companion. A heuristic approach to removing these is to subtract some suitably defined minimal flux from the lightcurve. The assumption is that this minimal flux is dominated by the constant contributions from the stars and the bright spot. We chose this flux as the threshold above which $85\%$ of the measurements are contained in the cumulative flux distribution. This choice gave a good estimate of the average quiescent flux.  We found that this leads to a decrease of the estimated $\dot{M}_t$ of at most 50\%, with most systems barely affected. As a cross-check, we estimated the contribution from the bright spot and the stars theoretically (Fig.~\ref{fig:cvmag}). The absolute $V$ magnitude of the companion and of the accretion-heated white dwarf are taken from \citet{2011ApJS..194...28K}. For the bright spot, we assumed accretion at the critical rate and computed the absolute $V$ magnitude assuming a temperature of 15\,000 K and a luminosity
\begin{equation}
L_{\rm BS} = \frac{GM\dot{M}_t}{2R_{\rm out}}\left[1-\frac{R_{\rm out}}{R_{\rm L1}}\right]. 
\end{equation}
In nearly all cases the theoretical contribution was fainter than the minimum flux set from the observations. The exceptions are FO Aql, IP Peg, SS Aur, and U Gem, for which the theoretical contribution was $\approx 2$ mag brighter (first two systems) or $\approx 1$ mag brighter (latter two), most likely because we assume $\dot{M}_t=\dot{M}_{\rm crit}^+$ when calculating the contribution from the bright spot, whereas $\dot{M}_t\ll \dot{M}_{\rm crit}^+$ (but it is not clear why other systems do not show the same trend). 

We conclude that the underestimation of $\dot{M}_t$ due to the model assumption and the incomplete lightcurve are roughly compensated by the overestimation of $\dot{M}_t$ due to contributions from the white dwarf, companion star, and bright spot. Hence, we decided not to include any systematic correction to the estimates presented here. The statistical error bars also provide a reasonable estimate of the systematic error on $\dot{M}_t$. In all cases, none of the modifications to our analysis that we explored led to results challenging the disk instability model: the dwarf novae were always consistently placed within the instability region of the parameter space.

\section{Conclusion}
We have estimated the mass transfer rate $\dot{M}_t$ in $\approx$130 CVs and found that their separation into stable (nova-likes) and unstable (dwarf novae) systems in the $(P_{\rm orb},\dot{M}_t)$ plane is fully consistent with the disk instability model (Fig.~\ref{fig:stability}). Although it was clear that nova-likes were going to have higher average accretion rates than dwarf novae, by virtue of their higher steady optical fluxes, it was not clear that the predicted instability line from the DIM would neatly separate the two sub-populations. We thus confirm a key prediction of the DIM that had not been systematically explored yet. We also confirm that irradiation heating of the disc plays a negligible role in the stability of CVs, unlike X-ray binaries where strong X-ray irradiation modifies the limit between stable and unstable systems \citep{2012MNRAS.424.1991C}. One of our motivations was to find systems that would challenge the DIM, much like SS\,Cyg has in the past before the question of its distance was settled (\S1). We have not uncovered such systems. The only system that clearly does not fit the DIM is AE Aqr,  well-known for its fast-spinning white dwarf propelling matter out of the system.

We focused on analysing a large sample of sources and very likely traded this with some accuracy in $\dot{M}_t$, especially for the binaries whose lightcurve contains many gaps. A more careful selection and analysis of the individual lightcurves, corrected for the various contributions to the optical flux and sources of extrinsic variability such as eclipses\footnote{Eclipsing systems are de facto removed from this study since we selected systems with $i\leq 80\degr$.}, together with a critical appraisal of the system parameters would undoubtedly improve the accuracy of some of the individual $\dot{M}_t$ that we derived in Tab.~\ref{table:DN}-\ref{table:NL}. These (time-consuming) refinements would allow to quantitatively investigate if/how the location of the binary in the $(P_{\rm orb},\dot{M}_t)$ plane reflects its variability pattern (sub-class type) and the deviations from the expected secular mass transfer rate. They would also allow an additional test of the DIM, albeit dependent on outburst physics, which is that systems closer to the critical mass transfer rate spend longer in outburst than unstable systems (the transientness diagram of \citealt{2012MNRAS.424.1991C}). At this stage, the uncertainties on system parameters and lightcurves make it difficult to robustly identify classes of variability with locations in the instability diagram.

The lack of knowledge on system parameters and the biases introduced by the irregular sampling of most of the lightcurves limit our ability to achieve this goal. The {\it Kepler} lightcurves of the dozen or so CVs that have been observed constitute the ideal dataset. We have not used them for lack of information on their system parameters. With those in hand, they would provide a sub-sample of unprecedented quality. In the future, similar regularly-sampled lightcurves may be expected as a by-product of searches for planetary transits, such as ESA's {\it Plato}, and synoptic surveys. For instance, the LSST will provide a huge sample of CV lightcurves, although the baseline cadence (1 visit every three nights) is not ideal. Even if the {\it Gaia} catalogue provides distances to these CVs, obtaining orbital periods and constraints on the system parameters from follow-up spectroscopy for this avalanche of data represents a daunting challenge for professional and amateur astronomers.

\begin{acknowledgement}
We thank the referee, John Thorstensen, for pointing out to us the 3D reddening map constructed by Green et al. 2018 (\url{http://argonaut.skymaps.info}), and Jorge Casares for correcting the mass ratio of SY Cnc. GD and JPL thank the hospitality of the KITP in Santa Barbara, where this work was initiated. GD and JPL acknowledge support from {\it Centre National d'Etudes Spatiales} (CNES). This research was supported  in part by the National Science Centre, Poland, grant 2015/19/B/ST9/01099 and by the National Science Foundation under Grant No. NSF PHY17-48958. This work has made use of the SVO Filter Profile Service (\url{http://svo2.cab.inta-csic.es/theory/fps/}) supported from the Spanish MINECO through grant AyA2014-55216, of the SIMBAD database and of the VizieR catalogue access tool, operated at CDS, Strasbourg, France. This work has made use of data from the European Space Agency (ESA) mission {\it Gaia} (\url{https://www.cosmos.esa.int/gaia}), processed by the {\it Gaia} Data Processing and Analysis Consortium (DPAC, \url{https://www.cosmos.esa.int/web/gaia/dpac/consortium}). Funding for the DPAC has been provided by national institutions, in particular the institutions participating in the {\it Gaia} Multilateral Agreement. We gratefully acknowledge the variable star observations from the AAVSO International Database contributed by observers worldwide. This research was made possible  through their patient monitoring of the activity of cataclysmic variables over decades.
\end{acknowledgement}

\bibliographystyle{aa}
\bibliography{dimcvs}

\begin{appendix}
\section{Online-only tables and figure}

\begin{figure*}
\begin{center}
\includegraphics[width=\linewidth]{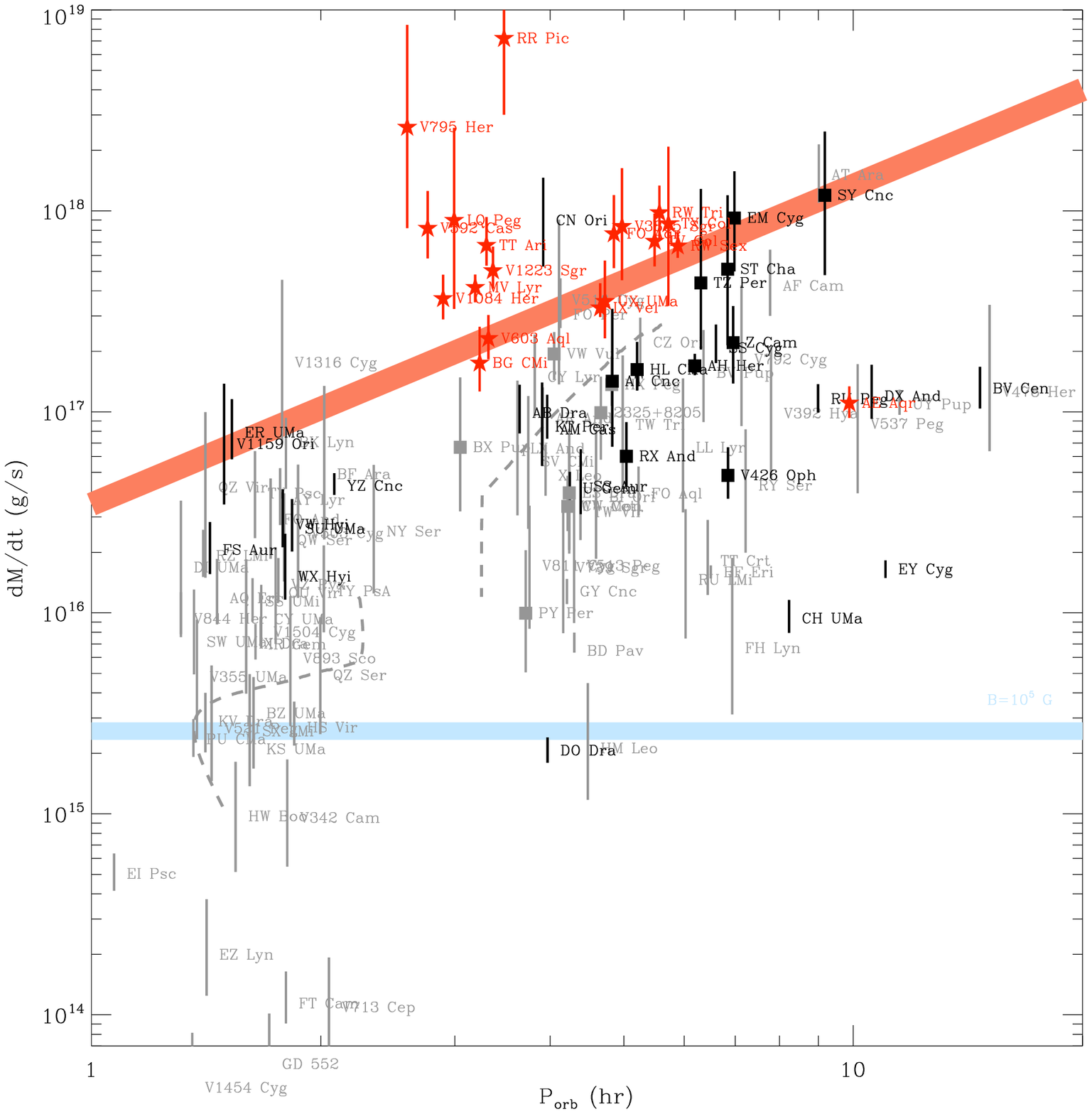} 
\caption{Same as Fig.~\ref{fig:stability} except the name of the system is indicated to the right of the estimated $\dot{M}_t$.}
\label{fig:stability_names}
\end{center}
\end{figure*}

\begin{table*}
\centering
\caption{Reconstructed mass transfer rate over one outburst cycle for various models. }
\label{table:model}
\begin{tabular}{cccccccccccccr}
\toprule\toprule
$M$& $\alpha_{\rm h}$ & $\alpha_{\rm c}$ & $R_{\rm circ}$ & $R_{\rm in}$ & $R_{\rm out}$  & $\dot{M}_t$ & $\dot{M}_U$ & $\dot{M}_B$ & $\dot{M}_V$ & $\dot{M}_R$ & $\dot{M}_I$  & $\mu$\\
{\small M$_\odot$} & & & {\small $10^{10}\rm\,cm$}& {\small $10^{8}\rm\,cm$} & {\small $10^{10}\rm\,cm$}  & {\small $10^{16}\rm\,g\,s^{-1}$ }& \multicolumn{5}{c}{{\small in units of $\dot{M}_t$}} & {\small $10^{30}\rm\,G\,cm^{3}$}\\
\midrule
0.6 & 0.2 & 0.04 & 1.00 & 8.67 & 1.90 &   10.0 & 0.92 & 0.89 & 0.87 & 0.86 & 0.86 & 0\\
0.6 & 0.2 & 0.04 & 1.00 & 8.67 & 1.97 &   1.00 & 1.03 & 0.97 & 0.91 & 0.87 & 0.84 & 0\\
0.6 & 0.2 & 0.04 & 1.00 & 8.67 & 2.00 &   0.10 & 1.23 & 1.13 & 1.02 & 0.94 & 0.85 & 0 \\
0.6 & 0.2 & 0.04 & 1.00 & 8.67 & 1.95 &   0.025 & 0.96 & 0.88 & 0.80 & 0.74 & 0.70 & 0 \\
0.6 & 0.2 & 0.04 & 2.00 & 8.67 & 3.92 &   10.0 & 0.96 & 0.92 & 0.87 & 0.85 & 0.83 & 0\\ 
0.6 & 0.2 & 0.04 & 2.00 & 8.67 & 3.98 &   1.00 & 0.96 & 0.89 & 0.81 & 0.74 & 0.67 & 0\\ 
0.6 & 0.2 & 0.04 & 2.00 & 8.67 & 3.98 &   0.10 & 0.98 & 0.89 & 0.78 & 0.68 & 0.59 & 0 \\ 
0.6 & 0.2 & 0.04 & 2.00 & 8.67 & 3.97 &   0.025 & 0.96 & 0.88 & 0.79 & 0.73 & 0.67 & 0 \\ 
1.2 & 0.2 & 0.04 & 2.00 & 3.85 & 6.03 &   10.0 & 1.23 & 1.17 & 1.10 & 1.06 & 1.04 & 0 \\ 
1.2 & 0.2 & 0.04 & 2.00 & 3.85 & 6.05 &   1.00 & 1.06 & 1.00 & 0.91 & 0.82 & 0.71 & 0\\ 
1.2 & 0.2 & 0.04 & 2.00 & 3.85 & 5.98 &   0.10 & 1.11 & 1.03 & 0.90 & 0.77 & 0.63 & 0\\ 
1.2 & 0.2 & 0.04 & 1.50 & 3.85 & 2.09 &   1.00 & 1.01 & 0.95 & 0.88 & 0.83 & 0.77 & 0\\ 
1.2 & 0.2 & 0.04 & 1.50 & 3.85 & 2.12 &   0.025 & 1.16 & 1.06 & 0.93 & 0.81 & 0.68 & 0\\ 
0.9 & 0.1 & 0.04 & 1.50 & 6.23 & 3.82 &   0.40 & 1.01 & 0.96 & 0.88 & 0.82 & 0.77 & 0 \\ 
0.9 & 0.16 & 0.04 & 1.50 & 6.23 & 3.84 &   0.40 & 1.03 & 0.96 & 0.86 & 0.77 & 0.68 & 0\\ 
0.9 & 0.10 & 0.03 & 1.50 & 6.23 & 3.86 &   1.00 & 1.07 & 1.01 & 0.92 & 0.85 & 0.78 & 0\\ 
0.9 & 0.10 & 0.03 & 1.50 & 6.23 & 3.83 &   0.05 & 1.04 & 0.95 & 0.84 & 0.74 & 0.65 & 0\\ 
0.9 & 0.12 & 0.03 & 1.00 & 6.23 & 1.98 &   0.10 & 1.12 & 1.04 & 0.93 & 0.85 & 0.77 &  0\\ 
0.6 & 0.23 & 0.03 & 1.00 & 8.67 & 1.99 &   1.00 & 0.94 & 0.88 & 0.83 & 0.81 & 0.79 & 0\\ 
0.6 & 0.23 & 0.03 & 1.00 & 8.67 & 1.94 &   0.10 & 0.96 & 0.88 & 0.81 & 0.75 & 0.69 & 0 \\ 
0.6 & 0.20 & 0.04 & 1.00 &  14.33 & 1.89 &  10.00 & 0.91 & 0.88 & 0.86 & 0.85 & 0.86 & 1\\ 
0.6 & 0.20 & 0.04 & 1.00 &  13.37 & 1.97 &   1.00 & 0.95 & 0.90 & 0.84 & 0.81 & 0.78 & 1\\ 
0.6 & 0.20 & 0.04 & 1.00 &  15.90 & 1.93 &   0.10 & 0.70 & 0.66 & 0.62 & 0.60 & 0.59 & 1\\ 
0.9 & 0.10 & 0.03 & 1.50 &  10.85 & 3.84 &   1.00 & 0.99 & 0.93 & 0.85 & 0.78 & 0.72 & 1 \\ 
0.9 & 0.10 & 0.03 & 1.50 &   7.07 & 3.83 &   0.05 & 0.99 & 0.91 & 0.80 & 0.71 & 0.63 & 0.1\\ 
1.2 & 0.20 & 0.04 & 2.00 &  14.34 & 5.96 &   0.10 & 0.68 & 0.64 & 0.57 & 0.51 & 0.45 & 1\\ 
\bottomrule
\end{tabular}
\end{table*}

\longtab[2]{
\begin{landscape}
\begin{longtable}{llccrrccccccl}
\caption{\label{table:DN}System parameters adopted for the cataclysmic variables in this work.}\\
\toprule\toprule
  object &      type &             $M_1$  &             $M_2$ &      $q$ &  $P_{\rm orb}$ &           $i$  & $\pi$  &  $A_V$ & $f$ & start & stop & $\dot{M}_t$ \\
            &               &     M$_\odot$ &    M$_\odot$ &             &                   hr & $\degr$ & mas &  mag  &      & \multicolumn{2}{c}{JD}    & g\,s$^{-1}$ \\
\midrule
\endfirsthead
\caption{continued.}\\
\toprule\toprule
  object &      type &             $M_1$  &             $M_2$ &      $q$ &  $P_{\rm orb}$ &           $i$  & $\pi$  &  $A_V$ & $f$ & start & stop & $\dot{M}_t$ \\
            &               &     M$_\odot$ &    M$_\odot$ &             &                   hr & $\degr$ & mas &  mag  &      & \multicolumn{2}{c}{JD}    & g\,s$^{-1}$ \\
\midrule
\endhead
\bottomrule
\endfoot
2325+8205 & ugz:     &       0.75  $\pm$       0.07  &       0.47  $\pm$       0.07  &       1.64  $\pm$       0.2  &  4.66 &      75.0  $\pm$       6  &  2.147 $\pm$ 0.038 &       0.277  $\pm$       0.040  &   0.19 &  54289 &  56521 & $( 0.6- 1.5)\times 10^{17}$ \\
AB Dra    & ug       &       0.80  $\pm$ {\em  0.16} &       0.34  $\pm$ {\em  0.07} &       2.35  $\pm$ {\em  0.7} &  3.65 &      40.0  $\pm$ {\em 10} &  2.442 $\pm$ 0.027 &       0.196  $\pm$       0.055  &   0.82 &  37130 &  56752 & $( 0.8- 1.4)\times 10^{17}$ \\
AF Cam    & ug       & {\em  0.75} $\pm$ {\em  0.15} & {\em  0.89} $\pm$ {\em  0.18} & {\em  0.84} $\pm$ {\em  0.2} &  7.78 &      45.0  $\pm$      10  &  1.065 $\pm$ 0.057 &       1.305  $\pm$       0.119  &   0.29 &  51806 &  56735 & $( 0.3- 0.6)\times 10^{18}$ \\
AH Her    & ugz      &       0.95  $\pm$       0.10  &       0.76  $\pm$       0.08  &       1.25  $\pm$       0.1  &  6.19 &      46.0  $\pm$       3  &  3.084 $\pm$ 0.030 &       0.060  $\pm$       0.023  &   0.84 &  40839 &  56751 & $( 1.5- 1.9)\times 10^{17}$ \\
AM Cas    & ug       &       0.55  $\pm$ {\em  0.11} & {\em  0.32} $\pm$ {\em  0.06} & {\em  1.71} $\pm$ {\em  0.5} &  3.96 &      18.0  $\pm$      18  &  2.303 $\pm$ 0.025 &       0.268  $\pm$       0.101  &   0.73 &  47422 &  56752 & $( 0.7- 1.2)\times 10^{17}$ \\
AQ Eri    & ugsu     & {\em  0.65} $\pm$ {\em  0.13} & {\em  0.08} $\pm$ {\em  0.02} &       7.75  $\pm$ {\em  2.2} &  1.46 &      55.0  $\pm$      10  &  2.658 $\pm$ 0.118 &       0.113  $\pm$       0.041  &   0.23 &  52797 &  56744 & $( 0.9- 1.9)\times 10^{16}$ \\
AR And    & ug       & {\em  0.75} $\pm$ {\em  0.15} & {\em  0.31} $\pm$ {\em  0.06} & {\em  2.40} $\pm$ {\em  0.7} &  3.91 &      40.0  $\pm$      10  &  2.278 $\pm$ 0.170 &       0.121  $\pm$       0.022  &   0.44 &  49907 &  56735 & $( 0.7- 1.3)\times 10^{17}$ \\
AT Ara    & ugss     &       0.53  $\pm$       0.14  &       0.42  $\pm$       0.10  &       1.27  $\pm$       0.1  &  9.01 &      38.0  $\pm$       5  &  1.019 $\pm$ 0.051 &       0.800  $\pm$       0.100  &   0.12 &  36286 &  56570 & $( 1.1- 2.1)\times 10^{18}$ \\
AT Cnc    & ugz      & {\em  0.75} $\pm$ {\em  0.15} & {\em  0.48} $\pm$ {\em  0.10} & {\em  1.57} $\pm$ {\em  0.4} &  4.83 & {\em 56.7} $\pm$ {\em 20} &  2.201 $\pm$ 0.047 &       0.088  $\pm$       0.043  &   0.55 &  46506 &  56752 & $( 0.7- 3.3)\times 10^{17}$ \\
AY Lyr    & ugsu     & {\em  0.75} $\pm$ {\em  0.15} & {\em  0.12} $\pm$ {\em  0.02} & {\em  6.16} $\pm$ {\em  1.7} &  1.77 &      41.0  $\pm$      10  &  2.216 $\pm$ 0.132 &       0.086  $\pm$       0.027  &   0.40 &  37450 &  56746 & $( 0.3- 0.5)\times 10^{17}$ \\
BD Pav    & ug       &       1.15  $\pm$       0.07  &       0.73  $\pm$       0.06  &       1.58  $\pm$       0.1  &  4.30 &      71.0  $\pm$       1  &  2.973 $\pm$ 0.033 &       0.000  $\pm$       0.100  &   0.22 &  50206 &  56591 & $( 0.6- 0.8)\times 10^{16}$ \\
BF Ara    & ugsu     & {\em  0.67} $\pm$ {\em  0.13} & {\em  0.15} $\pm$ {\em  0.03} &       4.35  $\pm$ {\em  1.2} &  2.02 & {\em 56.7} $\pm$ {\em 20} &  2.030 $\pm$ 0.226 &       0.800  $\pm$       0.100  &   0.13 &  48991 &  56739 & $( 0.2- 1.3)\times 10^{17}$ \\
BF Eri    & ug       &       1.28  $\pm$       0.05  &       0.52  $\pm$       0.01  &       2.44  $\pm$       0.1  &  6.50 &      40.0  $\pm$       1  &  1.786 $\pm$ 0.032 &       0.112  $\pm$       0.023  &   0.37 &  51413 &  56745 & $( 1.5- 1.7)\times 10^{16}$ \\
BI Ori    & ug       & {\em  0.75} $\pm$ {\em  0.15} & {\em  0.44} $\pm$ {\em  0.09} & {\em  1.71} $\pm$ {\em  0.5} &  4.60 & {\em 56.7} $\pm$ {\em 20} &  1.415 $\pm$ 0.074 &       0.125  $\pm$       0.027  &   0.37 &  50318 &  56741 & $( 0.2- 0.8)\times 10^{17}$ \\
BV Cen    & ug       &       1.24  $\pm$       0.22  &       1.10  $\pm$       0.19  &       1.12  $\pm$       0.3  & 14.67 &      53.0  $\pm$       4  &  2.728 $\pm$ 0.040 &       0.310  $\pm$       0.100  &   0.68 &  34864 &  51864 & $( 1.0- 1.7)\times 10^{17}$ \\
BV Pup    & ugss     & {\em  0.75} $\pm$ {\em  0.15} & {\em  0.70} $\pm$ {\em  0.14} & {\em  1.07} $\pm$ {\em  0.3} &  6.36 &      65.0  $\pm$      10  &  1.792 $\pm$ 0.027 &       0.119  $\pm$       0.019  &   0.35 &  43349 &  56743 & $( 0.9- 2.6)\times 10^{17}$ \\
BX Pup    & ugz      & {\em  0.75} $\pm$ {\em  0.15} & {\em  0.20} $\pm$ {\em  0.04} & {\em  3.75} $\pm$ {\em  1.1} &  3.05 & {\em 56.7} $\pm$ {\em 20} &  1.325 $\pm$ 0.041 &       0.138  $\pm$       0.048  &   0.20 &  43620 &  56751 & $( 0.3- 1.5)\times 10^{17}$ \\
BZ UMa    & ugsu     & {\em  0.76} $\pm$ {\em  0.15} & {\em  0.11} $\pm$ {\em  0.02} &       7.19  $\pm$ {\em  2.0} &  1.63 &      57.0  $\pm$      10  &  6.557 $\pm$ 0.064 &       0.112  $\pm$       0.019  &   0.40 &  48844 &  56739 & $( 2.4- 4.8)\times 10^{15}$ \\
CH UMa    & ug       &       1.26  $\pm$       0.18  &       0.96  $\pm$       0.01  &       1.31  $\pm$       0.1  &  8.24 &      21.0  $\pm$       4  &  2.627 $\pm$ 0.021 &       0.154  $\pm$       0.028  &   0.58 &  45684 &  56753 & $( 0.8- 1.2)\times 10^{16}$ \\
CN Ori    & ug       &       0.74  $\pm$       0.10  &       0.49  $\pm$       0.08  &       1.51  $\pm$       0.1  &  3.92 &      67.0  $\pm$       3  &  2.844 $\pm$ 0.044 &       0.478  $\pm$       0.291  &   0.63 &  41350 &  56751 & $( 0.5- 1.5)\times 10^{18}$ \\
CW Mon    & ug       & {\em  0.75} $\pm$ {\em  0.15} & {\em  0.37} $\pm$ {\em  0.07} & {\em  2.01} $\pm$ {\em  0.6} &  4.24 &      67.0  $\pm$      10  &  2.916 $\pm$ 0.075 &       0.346  $\pm$       0.162  &   0.29 &  53816 &  56736 & $( 0.2- 0.6)\times 10^{17}$ \\
CY Lyr    & ug       & {\em  0.75} $\pm$ {\em  0.15} & {\em  0.30} $\pm$ {\em  0.06} & {\em  2.54} $\pm$ {\em  0.7} &  3.82 &      60.0  $\pm$      10  &  2.043 $\pm$ 0.041 &       0.202  $\pm$       0.056  &   0.45 &  41635 &  56744 & $( 1.0- 2.4)\times 10^{17}$ \\
CY UMa    & ugsu     & {\em  0.69} $\pm$ {\em  0.14} & {\em  0.11} $\pm$ {\em  0.02} &       6.29  $\pm$ {\em  1.8} &  1.67 &      55.0  $\pm$      10  &  3.184 $\pm$ 0.135 &       0.052  $\pm$       0.014  &   0.23 &  49614 &  56745 & $( 0.7- 1.4)\times 10^{16}$ \\
CZ Ori    & ug       &       0.55  $\pm$ {\em  0.11} & {\em  0.55} $\pm$ {\em  0.11} & {\em  1.00} $\pm$ {\em  0.3} &  5.25 &      18.0  $\pm$      10  &  2.022 $\pm$ 0.109 &       0.558  $\pm$       0.084  &   0.36 &  37505 &  56752 & $( 1.7- 2.9)\times 10^{17}$ \\
DI UMa    & ugsu     & {\em  0.75} $\pm$ {\em  0.15} & {\em  0.05} $\pm$ {\em  0.01} & {\em 14.08} $\pm$ {\em  4.0} &  1.31 & {\em 56.7} $\pm$ {\em 20} &  1.456 $\pm$ 0.078 &       0.073  $\pm$       0.024  &   0.24 &  49251 &  56750 & $( 0.8- 3.6)\times 10^{16}$ \\
DO Dra    & ug/dq:   &       0.83  $\pm$       0.10  &       0.38  $\pm$       0.01  &       2.20  $\pm$       0.2  &  3.97 &      45.0  $\pm$       4  &  5.056 $\pm$ 0.030 &       0.055  $\pm$       0.026  &   0.61 &  50218 &  56740 & $( 1.8- 2.4)\times 10^{15}$ \\
DX And    & ug       & {\em  0.75} $\pm$ {\em  0.07} & {\em  0.50} $\pm$ {\em  0.05} &       1.50  $\pm$       0.2  & 10.57 &      45.0  $\pm$      12  &  1.669 $\pm$ 0.025 &       0.266  $\pm$       0.032  &   0.60 &  45493 &  56714 & $( 0.9- 1.7)\times 10^{17}$ \\
EI Psc    & ugsu     &       0.65  $\pm$       0.11  &       0.13  $\pm$       0.03  &       5.26  $\pm$       0.6  &  1.07 &      50.0  $\pm$       5  &  6.552 $\pm$ 0.067 &       0.177  $\pm$       0.037  &   0.28 &  52218 &  56662 & $( 0.4- 0.6)\times 10^{15}$ \\
EM Cyg    & ugz      &       1.00  $\pm$       0.06  &       0.77  $\pm$       0.08  &       1.30  $\pm$       0.1  &  6.98 &      67.0  $\pm$       2  &  1.834 $\pm$ 0.299 &       0.116  $\pm$       0.040  &   0.86 &  44151 &  56750 & $( 0.5- 1.6)\times 10^{18}$ \\
ER UMa    & ugsu     & {\em  0.73} $\pm$ {\em  0.16} &       0.10  $\pm$       0.02  &       6.95  $\pm$       0.2  &  1.53 &      45.0  $\pm$      10  &  2.676 $\pm$ 0.046 &       0.078  $\pm$       0.018  &   0.62 &  49249 &  56749 & $( 0.6- 1.2)\times 10^{17}$ \\
ES Dra    & ugsu/ugz &       0.58  $\pm$ {\em  0.12} & {\em  0.37} $\pm$ {\em  0.07} & {\em  1.55} $\pm$ {\em  0.4} &  4.24 & {\em 56.7} $\pm$ {\em 20} &  1.480 $\pm$ 0.030 &       0.060  $\pm$       0.021  &   0.40 &  49865 &  56746 & $( 0.2- 0.8)\times 10^{17}$ \\
EY Cyg    & ugss     &       1.10  $\pm$       0.09  &       0.49  $\pm$       0.09  &       2.27  $\pm$       0.1  & 11.02 &      14.0  $\pm$       1  &  1.544 $\pm$ 0.020 &       0.111  $\pm$       0.047  &   0.57 &  42126 &  56750 & $( 1.5- 1.8)\times 10^{16}$ \\
EZ Lyn    & ugsu/wz  & {\em  0.75} $\pm$ {\em  0.15} & {\em  0.08} $\pm$ {\em  0.02} & {\em  9.74} $\pm$ {\em  2.8} &  1.42 & {\em 56.7} $\pm$ {\em 20} &  6.866 $\pm$ 0.153 &       0.050  $\pm$       0.025  &   0.31 &  53800 &  56749 & $( 1.2- 3.8)\times 10^{14}$ \\
FH Lyn    & ug::     & {\em  0.75} $\pm$ {\em  0.09} & {\em  0.44} $\pm$ {\em  0.06} &       1.70  $\pm$       0.3  &  6.94 & {\em 56.7} $\pm$ {\em 20} &  0.565 $\pm$ 0.137 &       0.140  $\pm$       0.025  &   0.15 &  54058 &  56729 & $( 0.3- 1.9)\times 10^{16}$ \\
FO And    & ugsu     & {\em  0.75} $\pm$ {\em  0.15} & {\em  0.12} $\pm$ {\em  0.02} &       6.50  $\pm$ {\em  1.8} &  1.72 &      55.0  $\pm$      10  &  1.669 $\pm$ 0.155 &       0.123  $\pm$       0.022  &   0.29 &  45232 &  56727 & $( 1.9- 4.7)\times 10^{16}$ \\
FO Aql    & ugss     &       0.55  $\pm$ {\em  0.11} & {\em  0.55} $\pm$ {\em  0.11} & {\em  1.01} $\pm$ {\em  0.3} &  5.22 &      18.0  $\pm$      10  &  2.047 $\pm$ 0.052 &       0.597  $\pm$       0.181  &   0.33 &  45849 &  56613 & $( 0.3- 0.5)\times 10^{17}$ \\
FO Per    & ug       &       0.40  $\pm$ {\em  0.08} & {\em  0.35} $\pm$ {\em  0.07} & {\em  1.13} $\pm$ {\em  0.3} &  4.13 &      18.0  $\pm$      18  &  1.713 $\pm$ 0.053 &       0.622  $\pm$       0.090  &   0.49 &  44759 &  56750 & $( 2.6- 4.6)\times 10^{17}$ \\
FS Aur    & ug       &       0.55  $\pm$ {\em  0.11} & {\em  0.08} $\pm$ {\em  0.02} & {\em  6.93} $\pm$ {\em  2.0} &  1.43 &      41.0  $\pm$      10  &  1.829 $\pm$ 0.070 &       0.387  $\pm$       0.082  &   0.50 &  53563 &  56735 & $( 1.6- 2.8)\times 10^{16}$ \\
FT Cam    & ug       & {\em  0.75} $\pm$ {\em  0.15} & {\em  0.13} $\pm$ {\em  0.03} & {\em  5.97} $\pm$ {\em  1.7} &  1.80 &      40.0  $\pm$      10  &  4.014 $\pm$ 0.086 &       0.107  $\pm$       0.122  &   0.19 &  53081 &  56333 & $( 0.9- 1.6)\times 10^{14}$ \\
GD 552    & wz:      & {\em  0.77} $\pm$ {\em  0.15} &       0.04  $\pm$       0.03  &      19.23  $\pm$ {\em 14.9} &  1.71 & {\em 56.7} $\pm$ {\em 20} & 12.347 $\pm$ 0.047 &       0.014  $\pm$       0.010  &   0.44 &  53723 &  56658 & $( 0.4- 1.0)\times 10^{14}$ \\
GY Cnc    & ug       &       0.99  $\pm$       0.12  &       0.38  $\pm$       0.06  &       2.60  $\pm$       0.2  &  4.21 &      77.3  $\pm$       1  &  3.789 $\pm$ 0.068 &       0.054  $\pm$       0.020  &   0.18 &  50907 &  56746 & $( 1.1- 1.5)\times 10^{16}$ \\
HL CMa    & ug/ugz   &       0.83  $\pm$       0.10  &       0.45  $\pm$       0.10  &       1.73  $\pm$ {\em  0.4} &  5.20 &      45.0  $\pm$      10  &  3.297 $\pm$ 0.040 &       0.094  $\pm$       0.035  &   0.63 &  44517 &  56751 & $( 1.3- 2.2)\times 10^{17}$ \\
HM Leo    & ug       & {\em  0.75} $\pm$ {\em  0.15} & {\em  0.42} $\pm$ {\em  0.08} & {\em  1.79} $\pm$ {\em  0.5} &  4.48 & {\em 56.7} $\pm$ {\em 20} &  1.983 $\pm$ 0.180 &       0.113  $\pm$       0.021  &   0.21 &  53625 &  56752 & $( 1.2- 4.5)\times 10^{15}$ \\
HS Vir    & ugsu     & {\em  0.68} $\pm$ {\em  0.14} & {\em  0.13} $\pm$ {\em  0.03} &       5.18  $\pm$ {\em  1.5} &  1.85 &      40.0  $\pm$      10  &  2.837 $\pm$ 0.056 &       0.064  $\pm$       0.015  &   0.31 &  52600 &  56711 & $( 2.2- 3.6)\times 10^{15}$ \\
HW Boo    & ug       & {\em  0.75} $\pm$ {\em  0.15} & {\em  0.10} $\pm$ {\em  0.02} & {\em  7.89} $\pm$ {\em  2.2} &  1.55 & {\em 56.7} $\pm$ {\em 20} &  2.524 $\pm$ 0.163 &       0.066  $\pm$       0.018  &   0.35 &  54125 &  56747 & $( 0.5- 1.8)\times 10^{15}$ \\
HX Peg    & ugz      & {\em  0.75} $\pm$ {\em  0.15} & {\em  0.48} $\pm$ {\em  0.10} & {\em  1.57} $\pm$ {\em  0.4} &  4.82 &      41.0  $\pm$      10  &  1.715 $\pm$ 0.056 &       0.151  $\pm$       0.016  &   0.49 &  50131 &  56684 & $( 1.1- 1.9)\times 10^{17}$ \\
IR Gem    & ugsu     & {\em  0.69} $\pm$ {\em  0.14} & {\em  0.11} $\pm$ {\em  0.02} &       6.49  $\pm$ {\em  1.8} &  1.64 &      15.0  $\pm$      10  &  3.758 $\pm$ 0.082 &       0.101  $\pm$       0.048  &   0.41 &  50346 &  56752 & $( 0.6- 0.9)\times 10^{16}$ \\
IX Dra    & ugsu     & {\em  1.40} $\pm$ {\em  0.28} & {\em  0.05} $\pm$ {\em  0.01} &      28.57  $\pm$ {\em  8.1} &  1.60 & {\em 56.7} $\pm$ {\em 20} &  1.266 $\pm$ 0.060 &       0.102  $\pm$       0.025  &   0.16 &  51802 &  56747 & $( 0.4- 1.8)\times 10^{16}$ \\
KS UMa    & ugsu     & {\em  0.94} $\pm$ {\em  0.19} & {\em  0.11} $\pm$ {\em  0.02} &       8.93  $\pm$ {\em  2.5} &  1.63 &      20.0  $\pm$      10  &  2.713 $\pm$ 0.112 &       0.084  $\pm$       0.019  &   0.30 &  52842 &  56726 & $( 1.7- 2.8)\times 10^{15}$ \\
KT Per    & ug       & {\em  0.75} $\pm$ {\em  0.15} & {\em  0.31} $\pm$ {\em  0.06} & {\em  2.41} $\pm$ {\em  0.7} &  3.90 &      60.0  $\pm$      10  &  4.128 $\pm$ 0.054 &       0.325  $\pm$       0.169  &   0.60 &  41745 &  56733 & $( 0.5- 1.4)\times 10^{17}$ \\
KV Dra    & ugsu     & {\em  0.71} $\pm$ {\em  0.14} & {\em  0.08} $\pm$ {\em  0.02} &       9.35  $\pm$ {\em  2.6} &  1.41 &      55.0  $\pm$      10  &  2.223 $\pm$ 0.060 &       0.047  $\pm$       0.016  &   0.26 &  51680 &  56723 & $( 2.0- 4.0)\times 10^{15}$ \\
LL Lyr    & ug       & {\em  0.75} $\pm$ {\em  0.15} & {\em  0.65} $\pm$ {\em  0.13} & {\em  1.15} $\pm$ {\em  0.3} &  5.98 & {\em 56.7} $\pm$ {\em 20} &  1.199 $\pm$ 0.078 &       0.084  $\pm$       0.016  &   0.14 &  37286 &  56637 & $( 0.3- 1.5)\times 10^{17}$ \\
LX And    & ug       & {\em  0.75} $\pm$ {\em  0.15} & {\em  0.26} $\pm$ {\em  0.05} & {\em  2.87} $\pm$ {\em  0.8} &  3.62 & {\em 56.7} $\pm$ {\em 20} &  1.855 $\pm$ 0.070 &       0.073  $\pm$       0.015  &   0.30 &  51512 &  56728 & $( 0.3- 1.4)\times 10^{17}$ \\
NY Ser    & ugsu     & {\em  0.81} $\pm$ {\em  0.16} & {\em  0.20} $\pm$ {\em  0.04} &       4.05  $\pm$ {\em  1.1} &  2.35 & {\em 56.7} $\pm$ {\em 20} &  1.294 $\pm$ 0.051 &       0.157  $\pm$       0.010  &   0.29 &  49830 &  56727 & $( 1.3- 5.4)\times 10^{16}$ \\
OU Vir    & ugsu     &       0.70  $\pm$       0.01  &       0.12  $\pm$       0.00  &       6.09  $\pm$       0.1  &  1.74 &      79.6  $\pm$       1  &  1.874 $\pm$ 0.396 &       0.112  $\pm$       0.033  &   0.17 &  53308 &  56448 & $( 0.7- 2.1)\times 10^{16}$ \\
PU CMa    & ugsu:    & {\em  0.46} $\pm$ {\em  0.09} & {\em  0.05} $\pm$ {\em  0.01} &       9.17  $\pm$ {\em  2.6} &  1.36 &      30.0  $\pm$      10  &  6.133 $\pm$ 0.031 &       0.032  $\pm$       0.013  &   0.24 &  53047 &  56348 & $( 1.9- 3.0)\times 10^{15}$ \\
PY Per    & ugz      & {\em  0.75} $\pm$ {\em  0.15} & {\em  0.28} $\pm$ {\em  0.06} & {\em  2.71} $\pm$ {\em  0.8} &  3.72 & {\em 56.7} $\pm$ {\em 20} &  1.990 $\pm$ 0.106 &       0.076  $\pm$       0.030  &   0.34 &  50407 &  56751 & $( 0.5- 2.0)\times 10^{16}$ \\
QW Ser    & ugsu     & {\em  0.85} $\pm$ {\em  0.17} & {\em  0.12} $\pm$ {\em  0.02} &       6.80  $\pm$ {\em  1.9} &  1.79 &      60.0  $\pm$      10  &  2.655 $\pm$ 0.262 &       0.072  $\pm$       0.035  &   0.22 &  52758 &  56718 & $( 1.4- 3.9)\times 10^{16}$ \\
QZ Ser    & ug       & {\em  0.75} $\pm$ {\em  0.15} & {\em  0.15} $\pm$ {\em  0.03} & {\em  4.95} $\pm$ {\em  1.4} &  2.00 & {\em 56.7} $\pm$ {\em 20} &  3.107 $\pm$ 0.095 &       0.195  $\pm$       0.054  &   0.29 &  52310 &  56556 & $( 0.2- 1.0)\times 10^{16}$ \\
QZ Vir    & ugsu/dq: & {\em  0.71} $\pm$ {\em  0.02} & {\em  0.08} $\pm$ {\em  0.00} &       9.26  $\pm$       0.3  &  1.41 &      65.0  $\pm$      19  &  7.814 $\pm$ 0.069 &       0.037  $\pm$       0.035  &   0.39 &  52633 &  56742 & $( 0.1- 1.0)\times 10^{17}$ \\
RU LMi    & ug       & {\em  0.75} $\pm$ {\em  0.15} & {\em  0.66} $\pm$ {\em  0.13} & {\em  1.14} $\pm$ {\em  0.3} &  6.02 & {\em 56.7} $\pm$ {\em 20} &  0.992 $\pm$ 0.086 &       0.039  $\pm$       0.005  &   0.21 &  50097 &  56746 & $( 0.7- 3.3)\times 10^{16}$ \\
RU Peg    & ugss     &       1.06  $\pm$       0.04  &       0.96  $\pm$       0.08  &       1.14  $\pm$       0.0  &  8.99 &      41.0  $\pm$       7  &  3.616 $\pm$ 0.045 &       0.112  $\pm$       0.032  &   0.75 &  38006 &  56685 & $( 1.0- 1.4)\times 10^{17}$ \\
RX And    & ugz      &       1.14  $\pm$       0.33  &       0.48  $\pm$       0.03  &       2.38  $\pm$       0.7  &  5.04 &      51.0  $\pm$       9  &  5.028 $\pm$ 0.051 &       0.044  $\pm$       0.024  &   0.75 &  26469 &  56741 & $( 0.4- 0.9)\times 10^{17}$ \\
RY Ser    & ug       & {\em  0.75} $\pm$ {\em  0.09} & {\em  0.62} $\pm$ {\em  0.07} &       1.20  $\pm$       0.2  &  7.22 & {\em 56.7} $\pm$ {\em 20} &  1.549 $\pm$ 0.054 &       0.849  $\pm$       0.055  &   0.12 &  50952 &  56743 & $( 0.2- 0.8)\times 10^{17}$ \\
RZ LMi    & ugsu     &       1.00  $\pm$ {\em  0.20} & {\em  0.07} $\pm$ {\em  0.01} & {\em 13.44} $\pm$ {\em  3.8} &  1.40 &      15.0  $\pm$      10  &  1.376 $\pm$ 0.084 &       0.072  $\pm$       0.018  &   0.40 &  50795 &  56751 & $( 1.5- 2.6)\times 10^{16}$ \\
SS Aur    & ug       &       1.08  $\pm$       0.40  &       0.39  $\pm$       0.02  &       2.80  $\pm$       1.0  &  4.39 &      40.0  $\pm$       7  &  3.849 $\pm$ 0.041 &       0.155  $\pm$       0.053  &   0.65 &  38010 &  56739 & $( 0.3- 0.7)\times 10^{17}$ \\
SS Cyg    & ugss     &       0.81  $\pm$       0.19  &       0.55  $\pm$       0.13  &       1.47  $\pm$       0.0  &  6.60 &      37.0  $\pm$       5  &  8.724 $\pm$ 0.049 &       0.017  $\pm$       0.046  &   1.00 &  55198 &  58119 & $( 1.7- 2.7)\times 10^{17}$ \\
SS UMi    & ugsu     & {\em  0.66} $\pm$ {\em  0.13} & {\em  0.10} $\pm$ {\em  0.02} &       6.33  $\pm$ {\em  1.8} &  1.63 &      37.0  $\pm$      10  &  1.887 $\pm$ 0.050 &       0.074  $\pm$       0.008  &   0.31 &  49967 &  56726 & $( 0.9- 1.5)\times 10^{16}$ \\
ST Cha    & ugz:     & {\em  0.75} $\pm$ {\em  0.15} & {\em  0.77} $\pm$ {\em  0.15} & {\em  0.98} $\pm$ {\em  0.3} &  6.84 & {\em 56.7} $\pm$ {\em 20} &  1.398 $\pm$ 0.025 &       0.500  $\pm$       0.100  &   0.51 &  54891 &  56751 & $( 0.2- 1.2)\times 10^{18}$ \\
SU UMa    & ugsu     & {\em  0.68} $\pm$ {\em  0.16} &       0.10  $\pm$       0.02  &       6.50  $\pm$       0.5  &  1.83 &      43.0  $\pm$      10  &  4.535 $\pm$ 0.029 &       0.044  $\pm$       0.046  &   0.75 &  40858 &  56747 & $( 2.0- 3.7)\times 10^{16}$ \\
SV CMi    & ug       & {\em  0.75} $\pm$ {\em  0.15} & {\em  0.28} $\pm$ {\em  0.06} & {\em  2.66} $\pm$ {\em  0.8} &  3.74 & {\em 56.7} $\pm$ {\em 20} &  2.196 $\pm$ 0.095 &       0.028  $\pm$       0.007  &   0.32 &  42662 &  56749 & $( 0.3- 1.2)\times 10^{17}$ \\
SW UMa    & ugsu/dq  &       0.71  $\pm$       0.22  &       0.10  $\pm$       0.01  &       7.10  $\pm$       2.0  &  1.36 &      45.0  $\pm$      18  &  6.148 $\pm$ 0.080 &       0.052  $\pm$       0.022  &   0.36 &  49586 &  56739 & $( 0.5- 1.3)\times 10^{16}$ \\
SX LMi    & ugsu     & {\em  0.67} $\pm$ {\em  0.13} & {\em  0.10} $\pm$ {\em  0.02} &       6.54  $\pm$ {\em  1.8} &  1.61 & {\em 56.7} $\pm$ {\em 20} &  3.081 $\pm$ 0.116 &       0.064  $\pm$       0.021  &   0.31 &  52706 &  56737 & $( 1.4- 5.0)\times 10^{15}$ \\
SY Cnc    & ugz      & {\em  0.75} $\pm$ {\em  0.06} & {\em  0.88} $\pm$ {\em  0.07} &       0.85  $\pm$       0.1  &  9.18 & {\em 56.7} $\pm$ {\em 20} &  2.232 $\pm$ 0.044 &       0.093  $\pm$       0.010  &   0.63 &  41562 &  56752 & $( 0.5- 2.5)\times 10^{18}$ \\
TT Crt    & ug       &       1.00  $\pm$ {\em  0.20} & {\em  0.71} $\pm$ {\em  0.14} & {\em  1.40} $\pm$ {\em  0.4} &  6.44 &      60.0  $\pm$      10  &  1.842 $\pm$ 0.067 &       0.062  $\pm$       0.025  &   0.29 &  52703 &  56700 & $( 1.2- 2.9)\times 10^{16}$ \\
TW Tri    & ug       & {\em  0.75} $\pm$ {\em  0.15} & {\em  0.51} $\pm$ {\em  0.10} & {\em  1.48} $\pm$ {\em  0.4} &  4.98 & {\em 56.7} $\pm$ {\em 20} &  1.379 $\pm$ 0.071 &       0.128  $\pm$       0.033  &   0.43 &  53432 &  56720 & $( 0.4- 1.9)\times 10^{17}$ \\
TW Vir    & ug       &       0.91  $\pm$       0.25  &       0.40  $\pm$       0.02  &       2.30  $\pm$       0.6  &  4.38 &      43.0  $\pm$      13  &  2.317 $\pm$ 0.117 &       0.061  $\pm$       0.016  &   0.23 &  35181 &  56748 & $( 0.2- 0.5)\times 10^{17}$ \\
TY PsA    & ugsu     & {\em  0.87} $\pm$ {\em  0.17} & {\em  0.15} $\pm$ {\em  0.03} &       5.62  $\pm$ {\em  1.6} &  2.02 &      65.0  $\pm$      10  &  5.431 $\pm$ 0.065 &       0.066  $\pm$       0.018  &   0.11 &  45311 &  56658 & $( 0.8- 2.2)\times 10^{16}$ \\
TY Psc    & ugsu     & {\em  0.70} $\pm$ {\em  0.14} & {\em  0.11} $\pm$ {\em  0.02} &       6.54  $\pm$ {\em  1.8} &  1.64 &      63.0  $\pm$      10  &  3.835 $\pm$ 0.097 &       0.139  $\pm$       0.035  &   0.34 &  52463 &  56711 & $( 0.2- 0.6)\times 10^{17}$ \\
TZ Per    & ugz      & {\em  0.75} $\pm$ {\em  0.15} & {\em  0.69} $\pm$ {\em  0.14} & {\em  1.08} $\pm$ {\em  0.3} &  6.31 & {\em 56.7} $\pm$ {\em 20} &  2.092 $\pm$ 0.025 &       0.350  $\pm$       0.260  &   0.79 &  37419 &  56752 & $( 0.2- 1.3)\times 10^{18}$ \\
U Gem     & ugss     &       1.17  $\pm$       0.15  &       0.44  $\pm$       0.06  &       2.80  $\pm$       0.1  &  4.25 &      69.0  $\pm$       2  & 10.712 $\pm$ 0.030 &       0.018  $\pm$       0.033  &   0.65 &  39035 &  56752 & $( 0.4- 0.5)\times 10^{17}$ \\
UY Pup    & ug       & {\em  0.75} $\pm$ {\em  0.05} & {\em  0.68} $\pm$ {\em  0.04} &       1.10  $\pm$       0.1  & 11.50 &      15.0  $\pm$      10  &  0.946 $\pm$ 0.040 &       0.164  $\pm$       0.020  &   0.28 &  43907 &  56739 & $( 1.0- 1.3)\times 10^{17}$ \\
V1159 Ori & ugsu     & {\em  0.72} $\pm$ {\em  0.16} &       0.11  $\pm$       0.02  &       6.82  $\pm$       0.2  &  1.49 &      35.0  $\pm$      10  &  2.844 $\pm$ 0.040 &       0.664  $\pm$       0.490  &   0.57 &  49605 &  56746 & $( 0.3- 1.4)\times 10^{17}$ \\
V1316 Cyg & ugsu:    & {\em  0.72} $\pm$ {\em  0.14} & {\em  0.12} $\pm$ {\em  0.02} &       5.85  $\pm$ {\em  1.7} &  1.78 & {\em 56.7} $\pm$ {\em 20} &  0.989 $\pm$ 0.023 &       1.129  $\pm$       0.093  &   0.29 &  52816 &  55617 & $( 0.7- 4.5)\times 10^{17}$ \\
V1454 Cyg & ugss     & {\em  0.75} $\pm$ {\em  0.15} & {\em  0.05} $\pm$ {\em  0.01} & {\em 14.99} $\pm$ {\em  4.2} &  1.36 & {\em 56.7} $\pm$ {\em 20} & 10.211 $\pm$ 1.790 &       0.027  $\pm$       0.026  &   0.28 &  53364 &  56138 & $( 0.3- 0.8)\times 10^{14}$ \\
V1504 Cyg & ugsu     & {\em  0.67} $\pm$ {\em  0.05} & {\em  0.11} $\pm$ {\em  0.01} &       6.10  $\pm$       0.6  &  1.67 &      30.0  $\pm$      10  &  1.897 $\pm$ 0.048 &       0.162  $\pm$       0.024  &   0.30 &  49504 &  56747 & $( 0.7- 1.0)\times 10^{16}$ \\
V342 Cam  & ugsu     & {\em  0.75} $\pm$ {\em  0.15} & {\em  0.13} $\pm$ {\em  0.03} & {\em  5.93} $\pm$ {\em  1.7} &  1.81 & {\em 56.7} $\pm$ {\em 20} &  3.525 $\pm$ 0.110 &       0.279  $\pm$       0.041  &   0.30 &  53470 &  56736 & $( 0.5- 1.9)\times 10^{15}$ \\
V355 UMa  & ugsu/wz  & {\em  0.75} $\pm$ {\em  0.15} & {\em  0.07} $\pm$ {\em  0.01} & {\em 10.96} $\pm$ {\em  3.1} &  1.38 & {\em 56.7} $\pm$ {\em 20} &  6.661 $\pm$ 0.086 &       0.048  $\pm$       0.011  &   0.22 &  54091 &  56741 & $( 0.2- 0.9)\times 10^{16}$ \\
V392 Hya  & ug:      & {\em  0.75} $\pm$ {\em  0.09} & {\em  0.42} $\pm$ {\em  0.05} &       1.80  $\pm$       0.3  &  7.80 & {\em 56.7} $\pm$ {\em 20} &  1.048 $\pm$ 0.049 &       0.117  $\pm$       0.029  &   0.18 &  51615 &  56421 & $( 0.5- 2.1)\times 10^{17}$ \\
V426 Oph  & ugz/dq:  &       0.90  $\pm$       0.19  &       0.70  $\pm$       0.14  &       1.29  $\pm$       0.1  &  6.85 &      59.0  $\pm$       6  &  5.196 $\pm$ 0.040 &       0.146  $\pm$       0.105  &   0.66 &  48418 &  56750 & $( 0.4- 0.7)\times 10^{17}$ \\
V478 Her  & ugsu     & {\em  0.75} $\pm$ {\em  0.15} & {\em  1.88} $\pm$ {\em  0.38} & {\em  0.40} $\pm$ {\em  0.1} & 15.10 & {\em 56.7} $\pm$ {\em 20} &  0.389 $\pm$ 0.059 &       0.166  $\pm$       0.024  &   0.46 &  53527 &  56706 & $( 0.6- 3.4)\times 10^{17}$ \\
V503 Cyg  & ugsu     & {\em  0.73} $\pm$ {\em  0.15} & {\em  0.13} $\pm$ {\em  0.03} &       5.46  $\pm$ {\em  1.5} &  1.87 & {\em 56.7} $\pm$ {\em 20} &  2.271 $\pm$ 0.088 &       0.207  $\pm$       0.071  &   0.40 &  50855 &  56634 & $( 1.2- 5.5)\times 10^{16}$ \\
V513 Peg  & ug       & {\em  0.75} $\pm$ {\em  0.15} & {\em  0.38} $\pm$ {\em  0.08} & {\em  1.95} $\pm$ {\em  0.6} &  4.30 & {\em 56.7} $\pm$ {\em 20} &  2.515 $\pm$ 0.062 &       0.157  $\pm$       0.023  &   0.16 &  54030 &  56658 & $( 0.9- 3.7)\times 10^{16}$ \\
V516 Cyg  & ugss     & {\em  0.75} $\pm$ {\em  0.15} & {\em  0.35} $\pm$ {\em  0.07} & {\em  2.14} $\pm$ {\em  0.6} &  4.11 & {\em 56.7} $\pm$ {\em 20} &  1.364 $\pm$ 0.042 &       0.542  $\pm$       0.233  &   0.43 &  46886 &  56740 & $( 0.1- 0.9)\times 10^{18}$ \\
V521 Peg  & ugsu     & {\em  0.75} $\pm$ {\em  0.15} & {\em  0.08} $\pm$ {\em  0.02} & {\em  9.32} $\pm$ {\em  2.6} &  1.44 & {\em 56.7} $\pm$ {\em 20} &  5.231 $\pm$ 0.087 &       0.071  $\pm$       0.035  &   0.10 &  53322 &  56574 & $( 1.5- 5.5)\times 10^{15}$ \\
V537 Peg  & -        & {\em  0.75} $\pm$ {\em  0.15} & {\em  1.21} $\pm$ {\em  0.24} & {\em  0.62} $\pm$ {\em  0.2} & 10.14 & {\em 56.7} $\pm$ {\em 20} &  1.356 $\pm$ 0.042 &       0.090  $\pm$       0.023  &   0.16 &  53659 &  56268 & $( 0.4- 1.7)\times 10^{17}$ \\
V630 Cas  & ug:      &       1.01  $\pm$       0.13  &       0.18  $\pm$       0.02  &       5.50  $\pm$       0.6  & 61.53 &      72.5  $\pm$       6  &  0.288 $\pm$ 0.053 &       0.460  $\pm$       0.026  &   0.67 &  53481 &  56700 & $( 0.3- 1.2)\times 10^{19}$ \\
V713 Cep  & ugsu:    & {\em  0.75} $\pm$ {\em  0.15} & {\em  0.16} $\pm$ {\em  0.03} & {\em  4.72} $\pm$ {\em  1.3} &  2.05 & {\em 56.7} $\pm$ {\em 20} &  2.900 $\pm$ 0.162 &       0.424  $\pm$       0.144  &   0.12 &  53896 &  56637 & $( 0.7- 1.9)\times 10^{14}$ \\
V729 Sgr  & ug       & {\em  0.75} $\pm$ {\em  0.15} & {\em  0.36} $\pm$ {\em  0.07} & {\em  2.09} $\pm$ {\em  0.6} &  4.16 & {\em 56.7} $\pm$ {\em 20} &  2.245 $\pm$ 0.041 &       0.256  $\pm$       0.013  &   0.14 &  52433 &  56376 & $( 0.8- 3.3)\times 10^{16}$ \\
V792 Cyg  & ugss     & {\em  0.75} $\pm$ {\em  0.15} & {\em  0.81} $\pm$ {\em  0.16} & {\em  0.93} $\pm$ {\em  0.3} &  7.13 & {\em 56.7} $\pm$ {\em 20} &  0.714 $\pm$ 0.050 &       0.222  $\pm$       0.031  &   0.25 &  53039 &  56607 & $( 0.8- 4.3)\times 10^{17}$ \\
V811 Cyg  & ugss     & {\em  0.75} $\pm$ {\em  0.15} & {\em  0.28} $\pm$ {\em  0.06} & {\em  2.64} $\pm$ {\em  0.7} &  3.76 & {\em 56.7} $\pm$ {\em 20} &  1.955 $\pm$ 0.046 &       0.234  $\pm$       0.056  &   0.13 &  42567 &  56554 & $( 0.8- 3.4)\times 10^{16}$ \\
V844 Her  & ugsu     & {\em  0.46} $\pm$ {\em  0.09} & {\em  0.05} $\pm$ {\em  0.01} &       8.70  $\pm$ {\em  2.5} &  1.31 &      40.0  $\pm$      10  &  3.406 $\pm$ 0.097 &       0.057  $\pm$       0.020  &   0.26 &  52661 &  56563 & $( 0.8- 1.3)\times 10^{16}$ \\
V893 Sco  & ug       &       0.89  $\pm$ {\em  0.18} &       0.17  $\pm$ {\em  0.04} &       5.30  $\pm$ {\em  1.5} &  1.82 &      75.0  $\pm$      10  &  8.056 $\pm$ 0.052 &       0.107  $\pm$       0.135  &   0.36 &  50904 &  56737 & $( 0.3- 1.2)\times 10^{16}$ \\
VW Hyi    & ugsu     &       0.67  $\pm$       0.22  &       0.11  $\pm$       0.03  &       6.80  $\pm$       0.2  &  1.78 &      60.0  $\pm$       3  & 18.531 $\pm$ 0.022 &       0.031  $\pm$       0.100  &   0.77 &  34597 &  56751 & $( 2.1- 4.1)\times 10^{16}$ \\
VW Vul    & ugsu/ugz &       0.35  $\pm$ {\em  0.07} & {\em  0.34} $\pm$ {\em  0.07} & {\em  1.03} $\pm$ {\em  0.3} &  4.05 &      41.0  $\pm$      10  &  1.847 $\pm$ 0.047 &       0.276  $\pm$       0.052  &   0.48 &  45908 &  56750 & $( 1.4- 2.5)\times 10^{17}$ \\
VZ Pyx    & ugsu     &       0.80  $\pm$ {\em  0.16} & {\em  0.12} $\pm$ {\em  0.02} &       6.80  $\pm$ {\em  1.9} &  1.76 &      35.0  $\pm$      10  &  3.976 $\pm$ 0.042 &       0.170  $\pm$       0.081  &   0.43 &  50959 &  56160 & $( 1.1- 1.9)\times 10^{16}$ \\
WW Cet    & ugz      &       0.83  $\pm$       0.16  &       0.41  $\pm$       0.07  &       2.00  $\pm$       0.2  &  4.22 &      54.0  $\pm$       9  &  4.588 $\pm$ 0.047 &       0.075  $\pm$       0.029  &   0.48 &  47886 &  56675 & $( 0.3- 0.5)\times 10^{17}$ \\
WX Hyi    & ugsu     &       0.90  $\pm$       0.30  &       0.16  $\pm$       0.05  &       5.50  $\pm$       1.5  &  1.80 &      40.0  $\pm$      10  &  4.273 $\pm$ 0.029 &       0.031  $\pm$       0.100  &   0.61 &  41066 &  56740 & $( 1.2- 2.5)\times 10^{16}$ \\
X Leo     & ug       &       1.03  $\pm$ {\em  0.21} & {\em  0.32} $\pm$ {\em  0.06} & {\em  3.23} $\pm$ {\em  0.9} &  3.95 &      41.0  $\pm$      10  &  2.520 $\pm$ 0.087 &       0.060  $\pm$       0.038  &   0.44 &  40763 &  56744 & $( 0.4- 0.7)\times 10^{17}$ \\
YZ Cnc    & ugsu     &       0.82  $\pm$       0.08  &       0.17  $\pm$ {\em  0.03} &       4.46  $\pm$ {\em  1.0} &  2.08 &      38.0  $\pm$       3  &  4.175 $\pm$ 0.046 &       0.083  $\pm$       0.051  &   0.61 &  42341 &  56752 & $( 0.4- 0.5)\times 10^{17}$ \\
Z Cam     & ugz      &       0.99  $\pm$       0.15  &       0.71  $\pm$       0.10  &       1.41  $\pm$       0.2  &  6.96 &      57.0  $\pm$      11  &  4.437 $\pm$ 0.040 &       0.042  $\pm$       0.016  &   0.92 &  25553 &  56753 & $( 1.4- 3.4)\times 10^{17}$ \\
\end{longtable}
\tablebib{System parameters from \citet{2017A&A...604A.107R}, \citet{2003A&A...404..301R} (v7.24), \citet{2012NewA...17..154G}, \cite{2012PASP..124..204P}, \citet{2011MNRAS.411.2695P}, \citet{2018arXiv180103555G}, \citet{2007NewA...12..446A},  \citet{2007ApJ...667.1139H}, \citet{2006ApJ...642.1029U}, \citet{2002AcA....52..429S}, \citet{1987A&AS...71..339V}, supplemented (values in italics) as necessary (see \S\ref{s:param}). Types are from \citet{2001PASP..113..764D} or \citet{2003A&A...404..301R}, with Z Cam types checked against \citet{2014JAVSO..42..177S}. Abbreviations are: ug, U Gem (generic dwarf nova); ss, SS Cyg subtype; su, SU UMa subtype; z, Z Cam subtype; dq, DQ Her subtype; wz, WZ Sge subtype, : indicates the type is uncertain. Two mass ratios from  \citet{2003A&A...404..301R} have been corrected: SY Cnc \citep{2009MNRAS.399.1534C}, QZ Vir \citep{2017PASJ...69...72I}.}
\end{landscape}}

\longtab[3]{
\begin{landscape}
\begin{longtable}{llccrrcccccclc}
\caption{\label{table:NL}Same as Tab.~\ref{table:DN} for nova-likes.}\\
\toprule\toprule
  object &      type &             $M_1$  &             $M_2$ &      $q$ &  $P_{\rm orb}$ &           $i$  & $\pi$  &  $A_V$ & $f$ & start & stop & $\dot{M}_t$  & $V$\\
            &               &     M$_\odot$ &    M$_\odot$ &             &                   hr & $\degr$ & mas &  mag  &      & \multicolumn{2}{c}{JD}    & g\,s$^{-1}$ & mag\\
\midrule
\endfirsthead
\caption{continued.}\\
\toprule\toprule
  object &      type &             $M_1$  &             $M_2$ &      $q$ &  $P_{\rm orb}$ &           $i$  & $\pi$  &  $A_V$ & $f$ & start & stop & $\dot{M}_t$ & $V$ \\
            &               &     M$_\odot$ &    M$_\odot$ &             &                   hr & $\degr$ & mas &  mag  &      & \multicolumn{2}{c}{JD}    & g\,s$^{-1}$ & mag \\
\midrule
\endhead
\bottomrule
\endfoot
AE Aqr    & dq       &       0.63  $\pm$       0.05  &       0.37  $\pm$       0.04  &       1.67  $\pm$       0.1  &  9.88 &      70.0  $\pm$       3  & 10.966 $\pm$ 0.055 &       0.027  $\pm$       0.029  &   0.56 &  32053 &  58068 & $( 0.9- 1.3)\times 10^{17}$ & 11.3 \\
BG CMi    & dq       &       0.80  $\pm$       0.20  &       0.38  $\pm$ {\em  0.08} & {\em  2.11} $\pm$ {\em  0.7} &  3.23 &      33.0  $\pm$      13  &  1.007 $\pm$ 0.055 &       0.069  $\pm$       0.017  &   0.02 &  46134 &  58157 & $( 1.3- 2.7)\times 10^{17}$ & 14.7 \\
BK Lyn    & nl/ug:   &       0.41  $\pm$       0.23  & {\em  0.13} $\pm$ {\em  0.03} &       5.00  $\pm$ {\em  3.0} &  1.80 &      32.0  $\pm$      12  &  1.980 $\pm$ 0.069 &       0.079  $\pm$       0.020  &   0.46 &  55935 &  56749 & $( 0.4- 0.9)\times 10^{17}$ & 14.6 \\
FO Aqr    & dq       & {\em  0.75} $\pm$ {\em  0.15} & {\em  0.48} $\pm$ {\em  0.10} & {\em  1.56} $\pm$ {\em  0.4} &  4.85 &      70.0  $\pm$       5  &  1.902 $\pm$ 0.051 &       0.091  $\pm$       0.023  &   0.03 &  50312 &  57313 & $( 0.5- 1.2)\times 10^{18}$ & 13.7 \\
IX Vel    & ux       &       0.82  $\pm$       0.14  &       0.53  $\pm$       0.09  &       1.54  $\pm$       0.1  &  4.65 &      57.0  $\pm$       2  & 11.042 $\pm$ 0.029 &       0.031  $\pm$       0.100  &   0.93 &  47644 &  51645 & $( 0.3- 0.4)\times 10^{18}$ &  9.6 \\
LQ Peg    & ux       & {\em  0.75} $\pm$ {\em  0.15} & {\em  0.20} $\pm$ {\em  0.04} & {\em  3.75} $\pm$ {\em  1.1} &  2.99 & {\em 56.7} $\pm$ {\em 20} &  0.923 $\pm$ 0.053 &       0.197  $\pm$       0.016  &   0.03 &  54321 &  57730 & $( 0.3- 2.6)\times 10^{18}$ & 14.7 \\
MV Lyr    & vy       &       0.73  $\pm$       0.10  &       0.30  $\pm$ {\em  0.06} &       2.30  $\pm$ {\em  0.6} &  3.19 &      10.0  $\pm$       3  &  2.002 $\pm$ 0.048 &       0.067  $\pm$       0.018  &   0.85 &  47879 &  49680 & $( 0.4- 0.5)\times 10^{18}$ & 12.4 \\
RR Pic    & nb       &       0.95  $\pm$ {\em  0.19} &       0.40  $\pm$ {\em  0.08} &       2.38  $\pm$ {\em  0.7} &  3.48 &      65.0  $\pm$      10  &  1.955 $\pm$ 0.030 &       0.062  $\pm$       0.100  &   0.48 &  47299 &  58165 & $( 0.3- 1.7)\times 10^{19}$ & 12.1 \\
RW Sex    & ux       &       0.90  $\pm$ {\em  0.05} &       0.67  $\pm$ {\em  0.04} &       1.35  $\pm$       0.1  &  5.88 &      34.0  $\pm$       6  &  4.228 $\pm$ 0.089 &       0.086  $\pm$       0.019  &   0.21 &  47449 &  55450 & $( 0.6- 0.8)\times 10^{18}$ & 10.5 \\
RW Tri    & ux       &       0.55  $\pm$       0.15  &       0.35  $\pm$       0.05  &       1.30  $\pm$       0.4  &  5.57 &      70.5  $\pm$       3  &  3.174 $\pm$ 0.048 &       0.286  $\pm$       0.046  &   0.17 &  48412 &  58139 & $( 0.7- 1.3)\times 10^{18}$ & 13.0 \\
TT Ari    & vy/dq:   &       0.90  $\pm$ {\em  0.18} &       0.20  $\pm$ {\em  0.04} &       4.50  $\pm$ {\em  1.3} &  3.30 &      30.0  $\pm$      10  &  3.884 $\pm$ 0.070 &       0.162  $\pm$       0.040  &   0.63 &  56198 &  58170 & $( 0.5- 0.9)\times 10^{18}$ & 10.8 \\
TV Col    & dq       &       0.75  $\pm$       0.15  &       0.56  $\pm$ {\em  0.11} & {\em  1.34} $\pm$ {\em  0.4} &  5.49 &      70.0  $\pm$       3  &  1.951 $\pm$ 0.018 &       0.155  $\pm$       0.100  &   0.31 &  48382 &  58170 & $( 0.5- 1.0)\times 10^{18}$ & 13.8 \\
TX Col    & dq       &       0.54  $\pm$       0.10  & {\em  0.62} $\pm$ {\em  0.12} & {\em  0.88} $\pm$ {\em  0.2} &  5.72 & {\em 56.7} $\pm$ {\em 20} &  1.084 $\pm$ 0.032 &       0.155  $\pm$       0.100  &   0.17 &  51418 &  52719 & $( 0.3- 2.1)\times 10^{18}$ & 14.5 \\
UX UMa    & ux       &       0.90  $\pm$       0.30  &       0.39  $\pm$       0.15  &       2.31  $\pm$       0.4  &  4.72 &      70.0  $\pm$       5  &  3.360 $\pm$ 0.023 &       0.045  $\pm$       0.009  &   0.73 &  47143 &  51144 & $( 0.2- 0.6)\times 10^{18}$ & 12.8 \\
V1084 Her & sw/dq:   & {\em  0.75} $\pm$ {\em  0.15} & {\em  0.20} $\pm$ {\em  0.04} & {\em  3.75} $\pm$ {\em  1.1} &  2.89 &      30.0  $\pm$      10  &  2.251 $\pm$ 0.030 &       0.059  $\pm$       0.014  &   0.17 &  56002 &  58126 & $( 0.3- 0.5)\times 10^{18}$ & 12.4 \\
V1223 Sgr & dq       &       0.93  $\pm$       0.12  &       0.33  $\pm$       0.08  & {\em  2.82} $\pm$ {\em  0.8} &  3.36 &      30.0  $\pm$      10  &  1.725 $\pm$ 0.047 &       0.307  $\pm$       0.037  &   0.32 &  49075 &  52076 & $( 0.4- 0.7)\times 10^{18}$ & 12.9 \\
V3885 Sgr & ux       &       0.70  $\pm$ {\em  0.14} &       0.47  $\pm$ {\em  0.10} &       1.47  $\pm$ {\em  0.4} &  4.97 &      65.0  $\pm$      10  &  7.537 $\pm$ 0.078 &       0.062  $\pm$       0.100  &   0.58 &  48008 &  49808 & $( 0.5- 1.6)\times 10^{18}$ & 10.4 \\
V592 Cas  & ux       & {\em  0.81} $\pm$ {\em  0.16} & {\em  0.20} $\pm$ {\em  0.04} &       4.03  $\pm$ {\em  1.1} &  2.76 &      40.0  $\pm$      10  &  2.198 $\pm$ 0.034 &       0.525  $\pm$       0.102  &   0.08 &  53998 &  58018 & $( 0.6- 1.3)\times 10^{18}$ & 12.6 \\
V603 Aql  & na/dq::  &       1.20  $\pm$       0.20  &       0.29  $\pm$       0.04  &       4.20  $\pm$       0.9  &  3.32 &      14.0  $\pm$       1  &  3.191 $\pm$ 0.069 &       0.221  $\pm$       0.123  &   0.53 &  31753 &  58067 & $( 1.8- 3.0)\times 10^{17}$ & 11.5 \\
V795 Her  & dq       & {\em  0.69} $\pm$ {\em  0.14} & {\em  0.20} $\pm$ {\em  0.04} &       3.45  $\pm$ {\em  1.0} &  2.60 & {\em 56.7} $\pm$ {\em 20} &  1.697 $\pm$ 0.039 &       0.071  $\pm$       0.017  &   0.05 &  50247 &  57938 & $( 0.8- 8.4)\times 10^{18}$ & 12.9 \\
\end{longtable}
\tablebib{Same references as Table \ref{table:DN}, including the typical $V$ magnitude. Abbreviations are:  nl, generic nova like; ux, UX UMa subtype; sw, SW Sex subtype; dq, DQ Her subtype; vy, VY Scl subtype; na and nb, fast and slow nova; : indicates the type is uncertain.}
\end{landscape}}

\end{appendix}

\end{document}